\documentclass[fleqn,usenatbib]{mnras}

\usepackage[T1]{fontenc}

\DeclareRobustCommand{\VAN}[3]{#2}
\let\VANthebibliography\thebibliography
\def\thebibliography{\DeclareRobustCommand{\VAN}[3]{##3}\VANthebibliography}

\usepackage{graphicx}	% Including figure files
\usepackage{amsmath}	% Advanced maths commands
\usepackage{amssymb}	% Extra maths symbols
\usepackage{xspace}

\newcommand{\kms}{\,km\,s$^{-1}$} % kilometres per second
\newcommand{\mJybeam}{\,mJy\,beam$^{-1}$} % mJy/beam
\newcommand{\uJybeam}{\,$\mu$Jy\,beam$^{-1}$} % uJy/beam
\newcommand{\Msunyr}{\,M$_\odot$\,yr$^{-1}$} % Msun/yr
\newcommand{\BD}{BD+43$^\circ$3654\xspace} % 

\title[EB27]{High-sensitivity radio study of the non-thermal stellar bow shock EB27}
\author[P. Benaglia et al.]{
Paula Benaglia,$^{1}$\thanks{E-mail: paula@iar-conicet.gov.ar}
Santiago del Palacio,$^{1}$
Christopher Hales$^{2}$
and Marcelo E. Colazo$^{3}$
\\
$^{1}$Instituto Argentino de Radioastronom\'{\i}a (CONICET;CICPBA;UNLP), C.C. No 5, 1894, Villa Elisa, Argentina\\
$^{2}$National Radio Astronomy Observatory, PO Box 0, Socorro, NM 87801, USA\\
$^{3}$Comisi\'{o}n Nacional de Actividades Espaciales, Paseo Col\'{o}n 751, 1063, CABA, Argentina
}

\date{Accepted Mar 1. Received Feb 16; in original form Dec 13, 2020}

\pubyear{2021}

\begin{document}
\label{firstpage}
\pagerange{\pageref{firstpage}--\pageref{lastpage}}
\maketitle

% Abstract of the paper
\begin{abstract}
%The abstract should briefly describe the aims, methods, and main results of the paper. It should be a single paragraph not more than 250 words. No references should appear in the abstract.
We present a deep radio-polarimetric observation of the stellar bow shock EB27 associated to the massive star BD+43$^\circ$3654. This is the only stellar bow shock confirmed to have non-thermal radio emission. We used the Jansky Very Large Array in S band (2--4~GHz) to test whether this synchrotron emission is polarised. The unprecedented sensitivity achieved allowed us to map even the fainter regions of the bow shock, revealing that the more diffuse emission is steeper and the bow shock brighter than previously reported.
No linear polarisation is detected in the bow shock above 0.5\%, although we detected polarised emission from two southern sources, probably extragalactic in nature. We modeled the intensity and morphology of the radio emission to better constrain the magnetic field and injected power in relativistic electrons. Finally, we derived a set of more precise parameters for the system EB27--BD+43$^\circ$3654 using \textit{Gaia} Early Data Release 3, including the spatial velocity. The new trajectory, back in time, intersects the core of the Cyg\,OB2 association.  
\end{abstract}

% Select between one and six entries 
\begin{keywords}
stars: massive, winds --- radiation mechanisms: non-thermal --- acceleration of particles 
\end{keywords}
 
%%%%%%%%%%%%%%%%% BODY OF PAPER %%%%%%%%%%%%%%%%%%

%=============================================================
%
\section{Introduction}\label{intro}
%
%=============================================================

When a runaway star with a powerful stellar wind travels through the interstellar medium (ISM) with a velocity  larger than the sound  speed of the ambient medium, 
it sweeps and heats the ambient dust and gas forming a bow-shaped feature that glows in infrared (IR) emission. These stellar bow shocks (BSs) started to be identified systematically by means of \textit{IRAS} data \citep[see][and references therein]{Noriega1997}.
Recently, \cite{Peri2012, Peri2015} presented the Extensive stellar BOw Shock Survey (E-BOSS) of about 70 objects, built mainly by searching all-sky \textit{WISE} images. Later, \cite{Kobulnicky2016} published the currently largest catalog with over 700 such structures by scrutinizing \textit{Spitzer} results. The shape and dynamics of stellar BSs have been addressed by several authors \citep[e.g.][]{Dyson1975, Wilkin1996, Meyer2016, Christie2016}. 

The presence of strong shocks allow for \textit{in situ} acceleration of relativistic particles \citep[e.g.][]{Bell1978}. Such particles can produce non-thermal emission from radio to gamma rays \citep{delValle2012, delValle2014, DelValle2018, delPalacio2018}. 
This seems to be the case of the object G70.7$+$1.2, that presents a  bow shock with non-thermal radio emission \citep{Kulkarni1992}, and where relativistic particles are most likely produced at a pulsar wind in a binary system, although its nature has  not been entirely settled \citep[see also][]{Cameron2007}.
Aside from compact objects, so far non-thermal emission has been unambiguously detected only in one stellar BS in the radio band \citep{Benaglia2010}. Studies at high energies have been unsuccessful at detecting the expected X-ray \citep{Toala2016, DeBecker2017, Toala2017} or $\gamma$-ray \citep{Schulz2014, HESS2018} emission from stellar BSs. The only exception is the possible association of two stellar BSs with \textit{Fermi} sources \citep{Sanchez-Ayaso2018}. Moreover, non-thermal radio emission from other BSs also remains elusive (C.S. Peri, private communication; Very Large Array -VLA- projects 13B-212, 16A-152).

In this context, obtaining further evidence of the non-thermal physics of stellar BSs is compelling. In particular, deep radio-polarimetric observations can provide us with additional information of the relativistic particle population and the magnetic field properties in the BSs. We report here on a polarimetric study of the surroundings of EB27, the bow shock from  which non-thermal emission was detected \citep{Benaglia2010}. Section~\ref{sec:EB27} presents what is known about this object and its proposed exciting star. In Sect.~\ref{sec:observations} the radio observations and related processing are described. The results are gathered in Sect.~\ref{sec:results}, analysed and discussed in Sect.~\ref{sec:disc}, and we close with some conclusions in Sect.~\ref{sec:conclusions}.

%======================================================
%
\section{The bow shock EB27 and the star  \BD}\label{sec:EB27}
%
%======================================================
% figure from file: CGPS-continuoRADec.fits
\begin{figure*}
    \centering
	\includegraphics[angle=-0, width=0.7\linewidth]{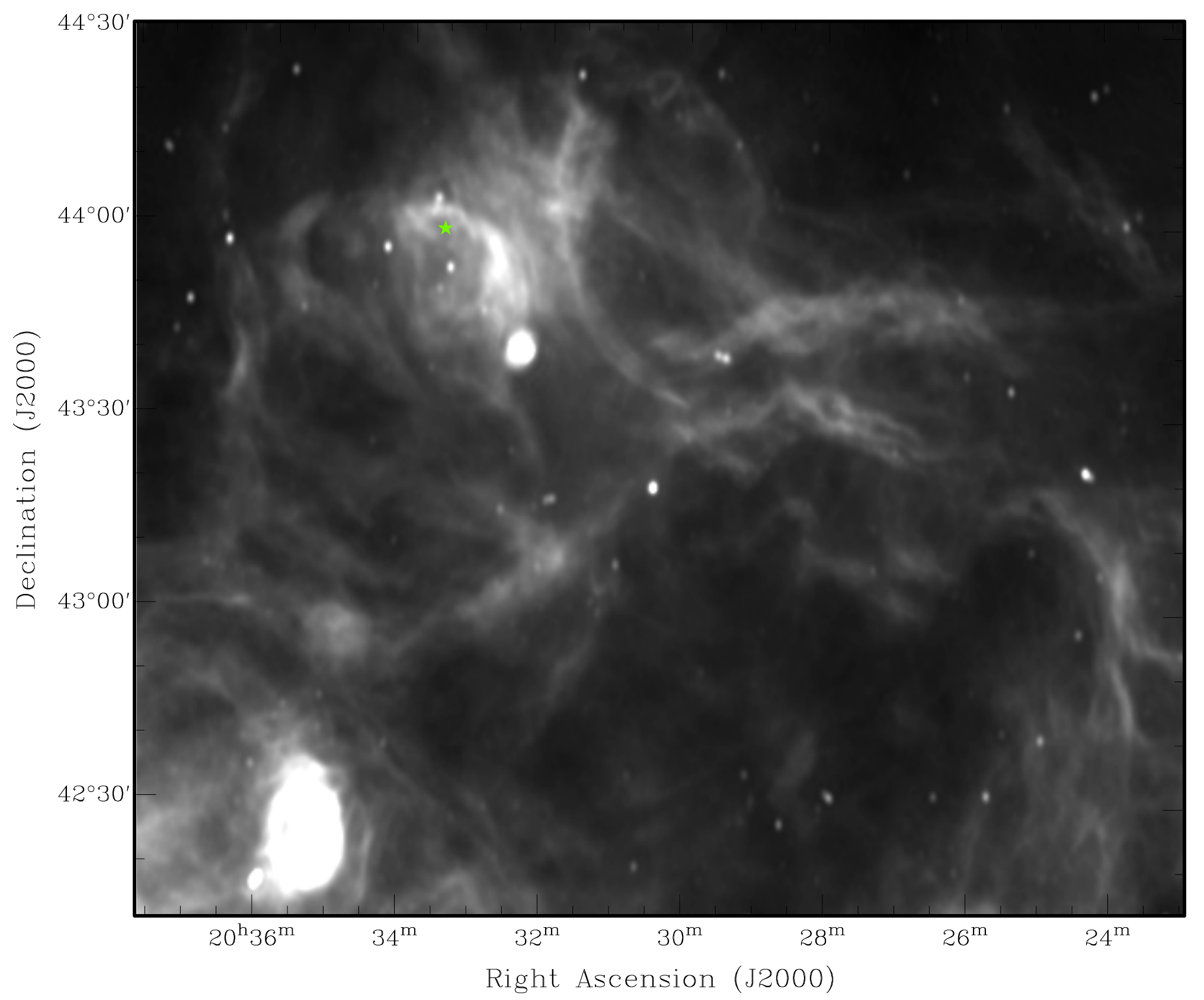}
    \caption{Continuum emission at 1420~MHz, as portrayed by the Canadian Galactic Plane Survey; data taken with the Dominion Radio Astrophysical Observatory, with angular resolution of $\sim1'  \times 1' (\cos \delta)^{-1}$, and typical rms of $\sim$0.3~mJy~beam$^{-1}$ \citep{CGPS2003}. The position of \BD is marked with a green star.}
    \label{fig:cgps}
\end{figure*}

The massive O4If star BD+43$^\circ$3654 has a strong, fast wind \citep[terminal velocity $v_\infty \approx 3000$\kms, e.g. ][]{Muijres2012} that, due to the supersonic motion of the star relative to the local ISM, produces a stellar BS. 
This BS was first identified by \cite{Comeron2007} using IR \textit{MSX} data, and it was later named EB27 by \cite{Peri2015}. The BS has an extension in the IR sky of 8'. The object lays almost at the Galactic plane  ($l,b =  82.45^\circ, +2.35^\circ$), in a region with bright radio emission at all angular scales, as testifies the image from the Canadian Galactic Plane Survey \citep{CGPS2003} that we reproduced in Fig.~\ref{fig:cgps}. EB27 was the first stellar BS to be detected at radio wavelengths, by \cite{Benaglia2010}. Their work presented VLA observations at 1.4 and 4.9 GHz, from which a mean negative spectral index alpha of $-0.5$ was measured in the BS emission (we adopt the convention $S \propto \nu^\alpha$). This spectrum was later corroborated with Giant Metrewave Radio Telescope observations at lower frequencies (610 and 1150~MHz) by \cite{Brookes2016}.
Their finding were indicative of non-thermal processes taking place at the BS, in particular relativistic particle acceleration -most likely due to diffusive shock acceleration at the shock- and synchrotron emission.

Under such conditions the relativistic electrons are expected to produce high-energy emission via inverse-Compton upscattering of ambient photons \citep[IR from the dust, optical from the star; e.g.][]{Benaglia2010}. 
One can characterise the non-thermal electron energy distribution and predict its high-energy emission by feeding theoretical models with the measured fluxes and spectral index in the radio band \citep[e.g.][]{delPalacio2018}. However, these predictions are really sensitive to even small uncertainties in the derived radio spectral index and other parameters related to rather unknown physical processes, such as the diffusion regime \citep{DelValle2018}.

Regarding the star, the latest results from the \textit{Gaia} mission\footnote{Early Data Release 3, https://gea.esac.esa.int/archive/} \citep[][]{Gaia2016b,Brown2020b} 
provided the most accurate astrometric measurements to date: $RA,  Dec {(\rm J2000)}$= 20:33:36.076, $+$43:59:07.38; parallax $\Pi = 0.582\pm0.012$~mas (thus $d = 1.72$~kpc); proper motions  $(\mu_\alpha \cos{\delta},\, \mu_\delta) = (-2.59\pm0.01,\, +0.73\pm0.01$)~mas\,yr$^{-1}$. The distance we obtained is significantly larger than the value of 1.32--1.4~kpc typically used in the literature % by \cite{Kobulnicky2017}. Also Toala used 1.4,  COmeron, etc.
for this object; however, this value is in great agreement with the value of $\approx 1.7$~kpc found more recently by other authors for Cyg OB2 \citep{Berlanas2019, Maiz2020}.

We used the former values to derive the spatial velocity of the star with respect to its local interstellar medium \citep[see e.g.][]{Comeron2007}. We first transformed the stellar proper motion to galactic coordinates, namely $(\mu_l \cos{b}, \mu_b) \approx (-0.96,\, 2.52)$~mas\,yr$^{-1}$. We then corrected these values by subtracting the local Galactic velocity field, calculated with the updated values of the Oort constants given by \cite{Bobylev2019}, getting $(\mu_l^\mathrm{G} \cos{b}, \, \mu_b^\mathrm{G}) \approx (-5.07, -0.92)$~mas\,yr$^{-1}$; we obtained $\Delta(\mu_l \cos{b},\, \mu_b) \approx (4.12, 3.43)$~mas\,yr$^{-1}$.

%Gaia Collaboration, A.G.A. Brown and et al. (2020b) Gaia early data release 3: summary of the contents and survey properties. A&A in prep.

\begin{figure}
    \centering
	\includegraphics[angle=-0, width=0.8\linewidth]{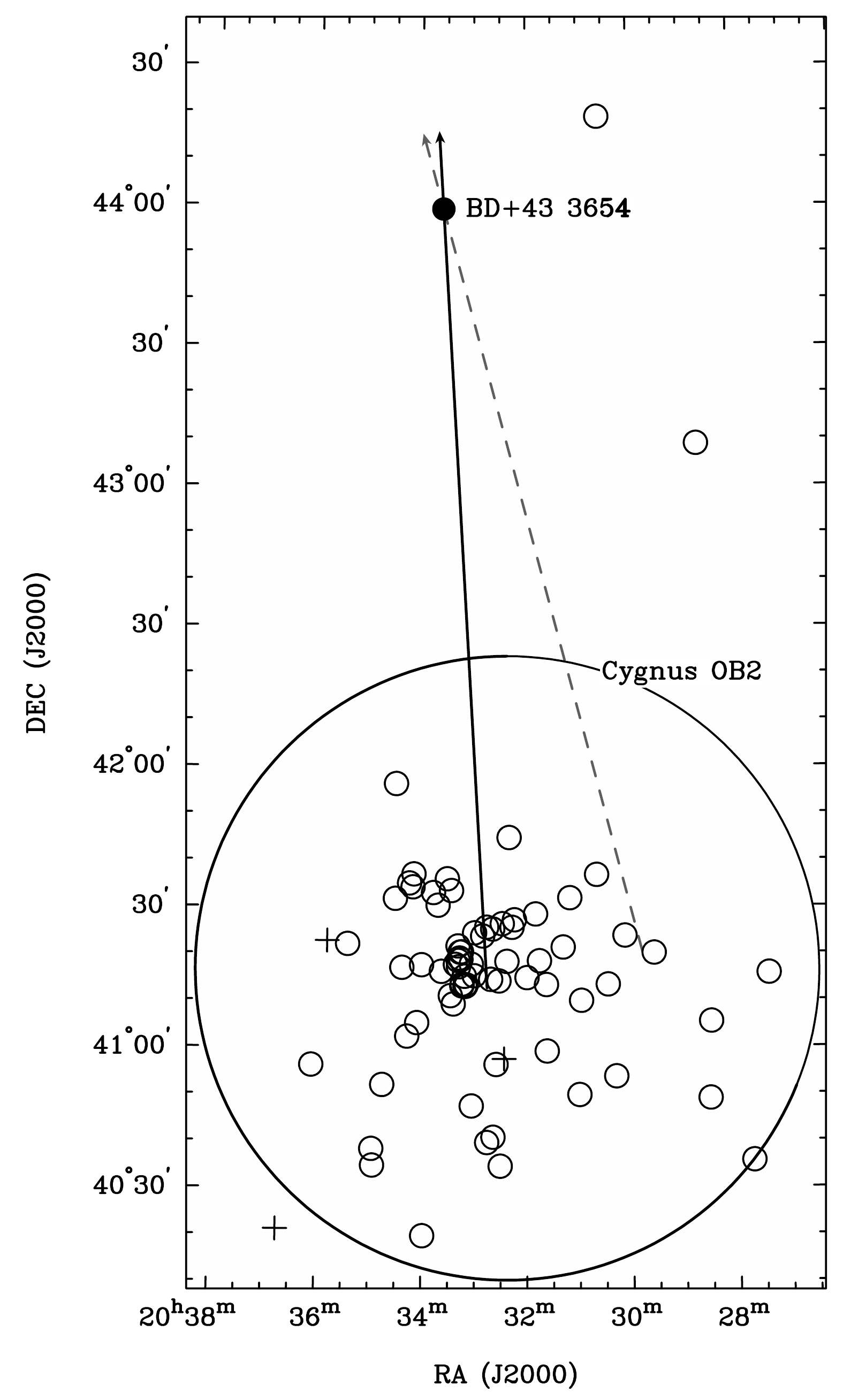}
    \caption{Direction of  motion (thicker arrow line) of the star BD+43$^\circ$3654 (filled circle). The line intersects the very central part of the association Cyg OB2, which is marked with the large circle \citep{Uyaniker2001}. The small crosses represent the Wolf-Rayet stars and the open circles, the O-type stars catalogued as in the literature \citep[see  Tables\,1, 2, and 6 from][and references therein]{Benaglia2020a}. The grey dashed line is the direction of motion by \citet{Comeron2007}.}
    \label{fig:bdtrajectory}
\end{figure}

Once we disaffected the proper motions from Galactic rotation, we could derive the tangential velocity $V_t \approx 43.6$~km~s$^{-1}$, and a direction of motion angle of $\approx -86.7\degr$ (measured North to East; see Figs.\,\ref{fig:bdtrajectory} and \ref{fig:I-50k}). As a reference, the values derived from \cite{Comeron2007} were $38.7$~km\,s$^{-1}$ and $-74.9\degr$, respectively. The new parameters are more consistent with the expectation of the stellar motion being aligned with the brightest region of the bow shock (see Fig.\,\ref{fig:I-50k}).

Finally, we projected the trajectory of BD+43$^\circ$3654 back in time using these new parameters. 
In principle, a runaway star moves with a constant velocity, as there is no friction with the ISM given that the stellar wind shields the star from its environment. 
In addition, we considered that the trajectory can be well-approximated as ballistic given that for the small scales involved (tens of pc) the effects of the Galactic potential would be neglectable in principle.
In Fig.\,\ref{fig:bdtrajectory} we  represent the path of the star, showing that it traverses the core  of Cyg\,OB2, crowded with OB systems. Assuming that the star moves at $V_t \approx 44$\kms, the time spent to travel that distance ($\approx 60$~pc) is $\sim$1.5~Myr. 
This value fits well with the estimated age of 1.6~Myr for \BD \citep{Comeron2007}.
According to \citet{Uyaniker2001}, the age of this OB association is 5~Myr. In conclusion, we provided additional support to the hypothesis that \BD is a runaway star from Cyg\,OB2, possibly  kicked off after close encounter(s) with other system(s).

%======================================================
%
\section{Observations and data reduction}\label{sec:observations}
%
%======================================================
%  

We observed the field of EB27 with a single Jansky Very Large Array (JVLA) pointing in D configuration at S band, on September 8, 2018. The time on source was 2.33~h. The flux calibrator used was 3C286, and the phase calibrator was J2052$+$3635. The polarisation calibrator was OQ208, an un-polarised source to obtain the $D$ terms. The observed band was centred at 3 GHz, and 2-GHz wide, divided in 16 spectral windows, of 64 channels each. We recorded full polarisation products (2\,MHz per channel, at 8 bit sampling, standard data rate of 2.4\,MB\,s$^{-1}$).

The data reduction was performed using the Common Astronomy Software Applications (CASA) package\footnote{https://casa.nrao.edu/}, mostly the 5.6.2 release. % \citep{casa}. 
The calibration processes included, besides standard steps, the application of the ionospheric correction routines which made use of Total Electron Content data. For the flux scale we used the \cite{Perley2017} scale. The position angle of 3C286 across the entire band was set according to \cite{Perley2013}. The last flagging was performed with \texttt{pieflag} \citep{Hales2014}.
% CAH: no need to say the following (the flux density of 3C286 comes from the model internally stored in CASA)
%The flux value of 3C286 resulted in 10.3$\pm$0.01~Jy~beam$^{-1}$.

The imaging steps were performed with CASA task \texttt{tclean}. Since EB27 is an extended source with many angular scales, we cleaned using multiple scales together with w-projection for each facet, to take into account the non-coplanarity of the baselines as a function of distance from the phase centre. We set a cell of 3$''$ and a maximum scale of 50~pixels. The cleaning algorithm included as well multi-term multi-frequency corrections to account for the frequency-dependent variation due to the wide fractional bandwidth.  A robust weighting of 0.25 provided the best combination of good angular resolution, low noise and contribution of large-scale emission. Out of 16 spectral windows, data from one of them (spectral window number 02) had to be flagged. 
The images were corrected for primary beam using \texttt{widebandpbcor} when needed. As the standard practice, the cleaning was handled playing with a large number of iterations % up to 50k
and a threshold, % down to 50muJy
in a combination such to grant convergence. 

The polarisation data processing consisted of performing rotation measure synthesis (RMS) of the linear polarisation. We used a python-based RMS code developed by M. R. Bell and collaborators,
\texttt{pyrmsynth}\footnote{https://github.com/mrbell/pyrmsynth, version 1.3.0.}. The routines request full Stokes cubes at individual spectral windows as input, and a parameters file containing information on the Faraday depth axis \citep[minimum, maximum and interval, see relevant equations in][]{Brentjens2005}. We consequently imaged the observed FoV at each spectral window and obtained the IQUV cubes. 
The routines  delivered an image of the polarised surface brightness $|P|$, related to the fractional polarisation (here called $D$) as $P = DI = Q + iU$, and 
performed RM cleaning of the spectra \citep[e.g.,][]{heald2009}.
The Faraday depth ($\phi$) interval probed, bounded by $\Delta \phi \approx \pi \lambda_{\rm min}^{-2}$, was $\pm10^{6}$~rad~m$^{-2}$. The resolution in Faraday space for our sample was 235~rad~m$^{-2}$.
 
%======================================================
%
\section{Results}\label{sec:results}
%
%======================================================

\subsection{Total intensity image and spectral index distribution} \label{sec:intensity}

\begin{figure*}
	\includegraphics[angle=-0, width=0.7\linewidth]{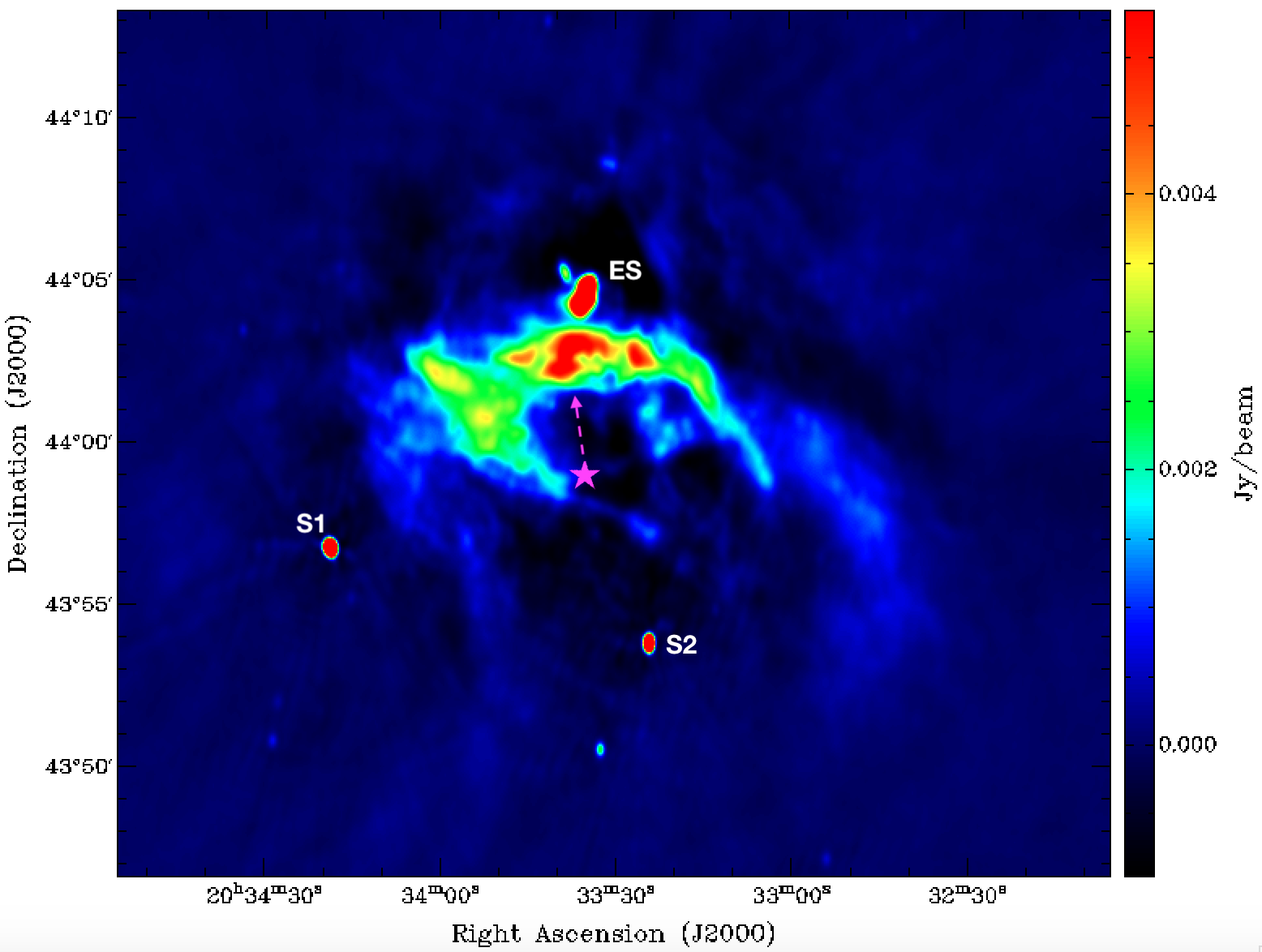}
    \caption{Continuum image at 3~GHz of the object EB27 (not corrected for primary beam). The sources ES, S1 and S2 are identified. The star symbol represents the position of \BD, and the dashed arrow its direction of motion.}
    \label{fig:I-50k}
\end{figure*}

Figure\,\ref{fig:I-50k} presents the I-Stokes image of the observed field.
The synthesised beam resulted in $20.2'' \times 12.5''$,  $PA=0.6$~deg. The rms, estimated over regions with no emission, is 0.1\mJybeam. At the position of the star \BD there is still some diffuse emission, at a mean level of 0.3\mJybeam (averaged over a 30$''$ size box). 
The thermal flux density expected from the star is lower than 0.1~mJy \citep[for the corresponding equations see for instance][]{Leitherer1995}. For this last calculation, we adopted a surface temperature of 40~kK, a mass-loss rate $\dot{M} \approx 5\times10^{-6}$\Msunyr \citep{Muijres2012},
%their Table~1, A-method: 4E-6. B-method: 3E-6, Vink+: 7E-6), 
a mean molecular weight of 1.5, and an rms ionic charge and mean number of electrons per ion of 1.
 
At S band, the radio brightness distribution observed with the JVLA mimics those of \cite{Benaglia2010} at L and C bands with the VLA (Fig.~\ref{fig:I-50k}). We detected up-close to the north, a source with a small-extension companion to the NE, dubbed ES (elliptical source) by \cite{Benaglia2010}. %The present deeper observations allowed to 
We also detected with very high signal-to-noise and dynamic range two unresolved sources to the SE and to the SW of EB27; we tagged these sources S1 and S2, respectively. 

The object here identified as ES corresponds to an H{\sc ii} region, reported by \citet{Lockman1989} as 82.454$+$2.369, a radio recombination line source. It was observed at 3~cm with 3$'$ angular resolution. Its radial velocity, $-9.7\pm2.4$~km~s$^{-1}$, implies a distance of $\sim$4~kpc.
% Rotation curve from Reid+2014

Gaussian fits to S1 and S2 yielded integrated flux densities of $S_{\rm S1} = 136\pm4$~mJy and $S_{\rm S2} = 96\pm2$~mJy, along the observed S band.
% image: I-12d-100mu-50k.pbcor.mir

According to the literature, S1 was previously detected at 327~MHz \citep[WSRTGP 2032$+$4346, 627$\pm130$~mJy,][]{Taylor1996}, at 365~MHz \citep[TXS\,2032$+$437, 499$\pm100$~mJy,][]{Douglas1996} and at 1.4~GHz \citep[NVSS\,J203418$+$435650, 115$\pm$3~mJy,][]{Condon1998}.
% S1: RA, Dec= 20:34:18.813, +43:56:49.823
And S2, at 327~MHz (WSRTGP 2031$+$4343, 157$\pm24$~mJy), and at 1.4~GHz (NVSS\,J20332$+$435353, 84$\pm$2~mJy).
% S2: RA, Dec= 20:33:24.280, +43:53:53.542
Both S1 and S2 sources show spectral indices favouring an extragalactic nature, probably AGN, from which we expect variable flux densities on years or even months. We therefore refrain to derive spectral indices from measurements years/decades apart and, in the next Subsection, we focus only on the behaviour at the 2-GHz width of the observed S-band.
 
The bibliographic information related to ES, S1 and S2, points at distinct sources, detached from EB27, in very different environments, and thus, all physically unrelated.

\begin{figure}
	\includegraphics[angle=-0, width=0.99\linewidth]{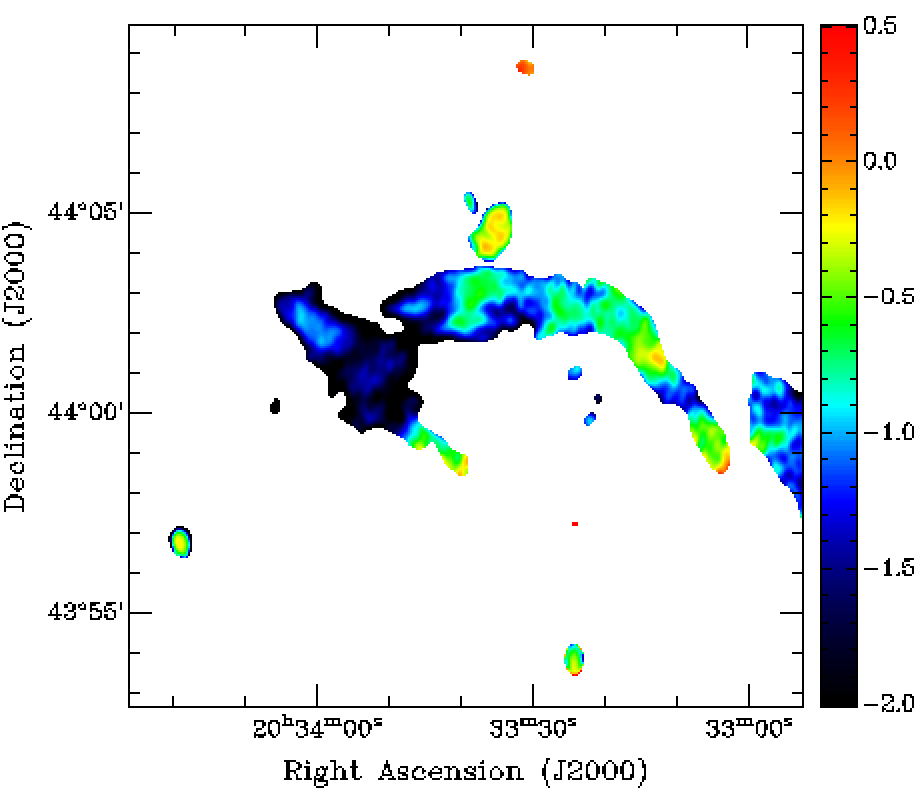}
	\includegraphics[angle=-0, width=0.99\linewidth]{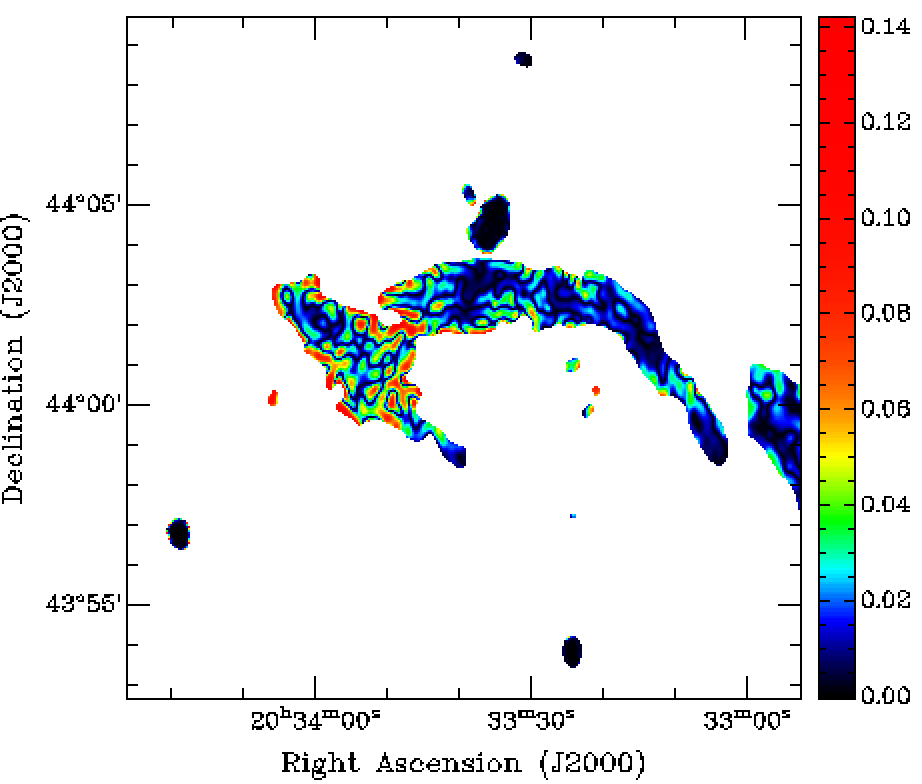}
    \caption{Upper panel: Spectral index $\alpha$. Lower panel: Error in spectral index. Mask used: pixels above 1.5~mJy~beam$^{-1}$.}
    \label{fig:alpha-y-error}
\end{figure}

Along with the total intensity image, \texttt{tclean} provides the spectral index map and its error. Figure\,\ref{fig:alpha-y-error} presents these results. The images are masked with $S_{\rm I} \geq 1.5$~mJy~beam$^{-1}$ to remove low signal-to-noise regions with unreliable spectral index determinations. 
% The spectral index images correspond to the primary  beam corrected I-image.

%------------------------------------------------
\subsection{Results per spectral window}

\begin{figure}
    \centering
    \includegraphics[width=0.99\linewidth]{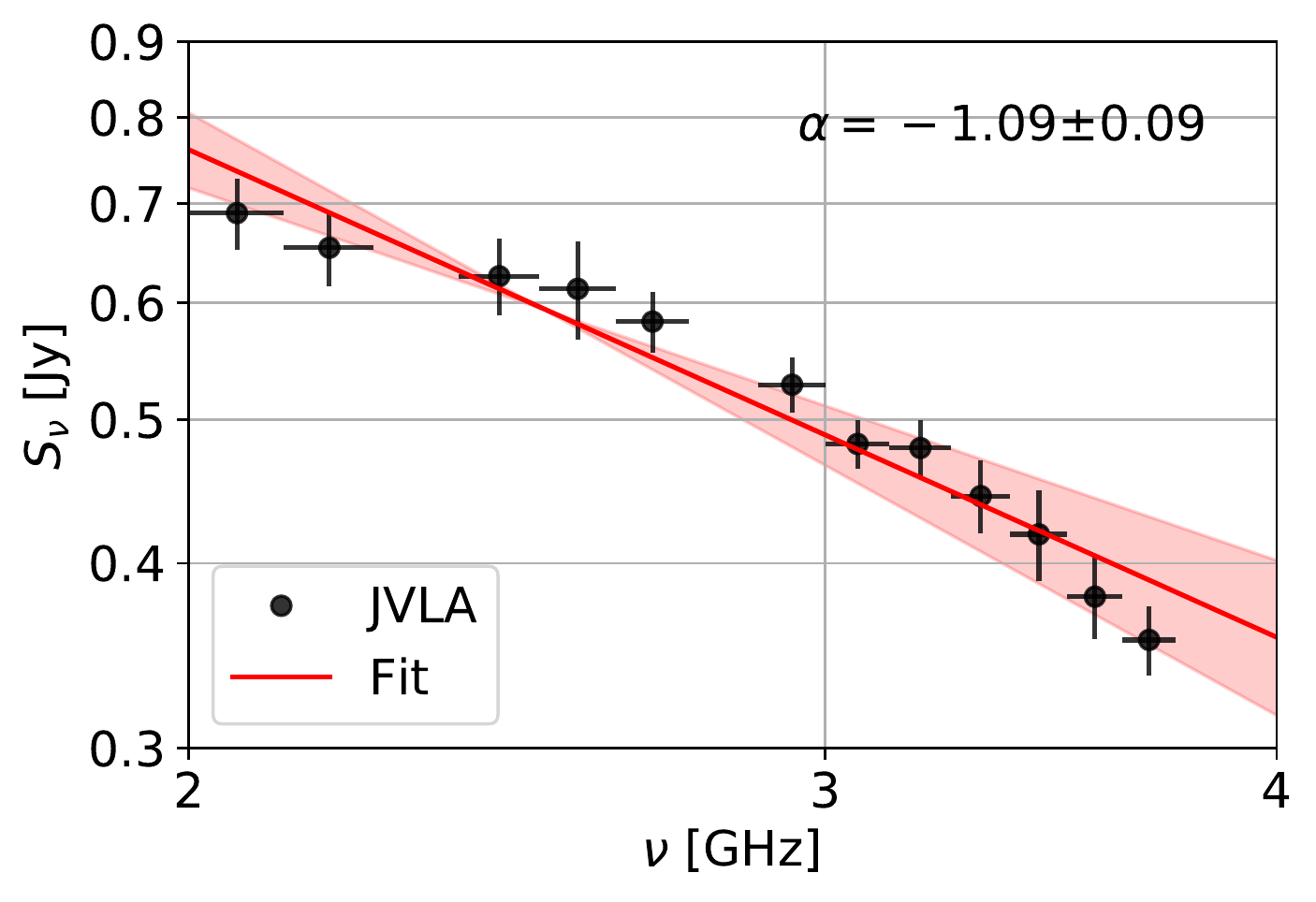}
    \includegraphics[width=0.99\linewidth]{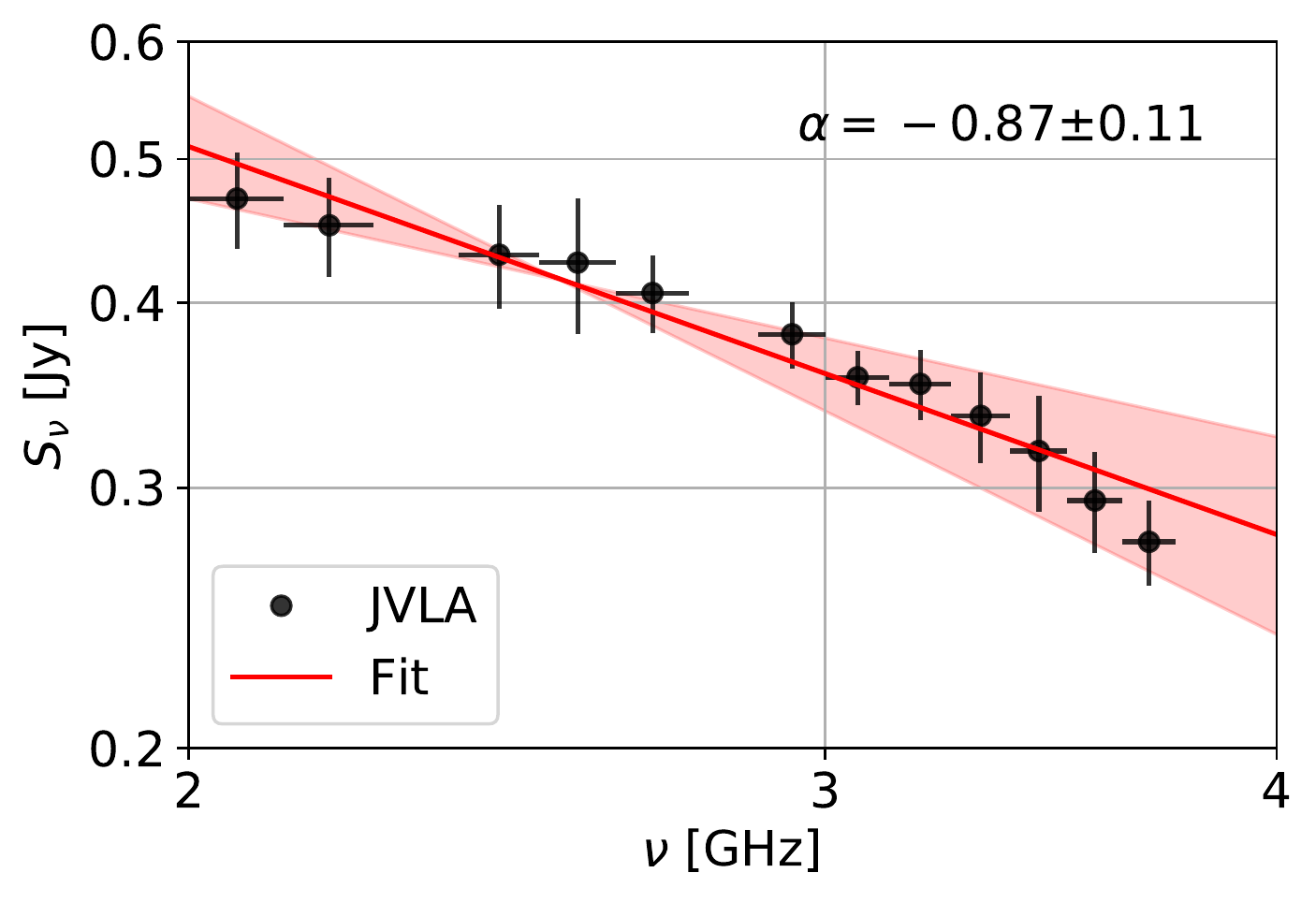}
    \caption{SED for the BS taken from contours $>$ 1.5\mJybeam (top) and $>$ 2.3\mJybeam (bottom). Filled lines represent 3-$\sigma$ uncertainties in the fit.}
    \label{fig:SED_BS}
\end{figure}

We built I-Stokes images per spectral window, and plotted the integrated flux of EB27 above two different intensity contours:
%(see Fig.~\ref{fig:SED_BS}): 
1.5 and 2.3~mJy~beam$^{-1}$. We assumed a 3\% uncertainty in the flux at each spectral window. We carried out a fit of the spectral index, and obtained $\alpha=-1.09\pm0.09$ and $\alpha=-0.87\pm0.11$, respectively (Fig.~\ref{fig:SED_BS}). This is consistent with the fainter regions being steeper, as seen in Fig.~\ref{fig:alpha-y-error}.
We decided to exclude from the spectral energy distribution (SED) the integrated fluxes reported in \citet{Benaglia2010}. This is because the present observations not only were much deeper, but also the current CASA routines to deal mainly with non-coplanarity effects, emission at very different scales, and extended emission far from the field centre, are in the state of the art and allow to recover much more flux density than those associated with VLA data back a dacade ago. 

\begin{figure}
    \centering
    \includegraphics[width=0.99\linewidth]{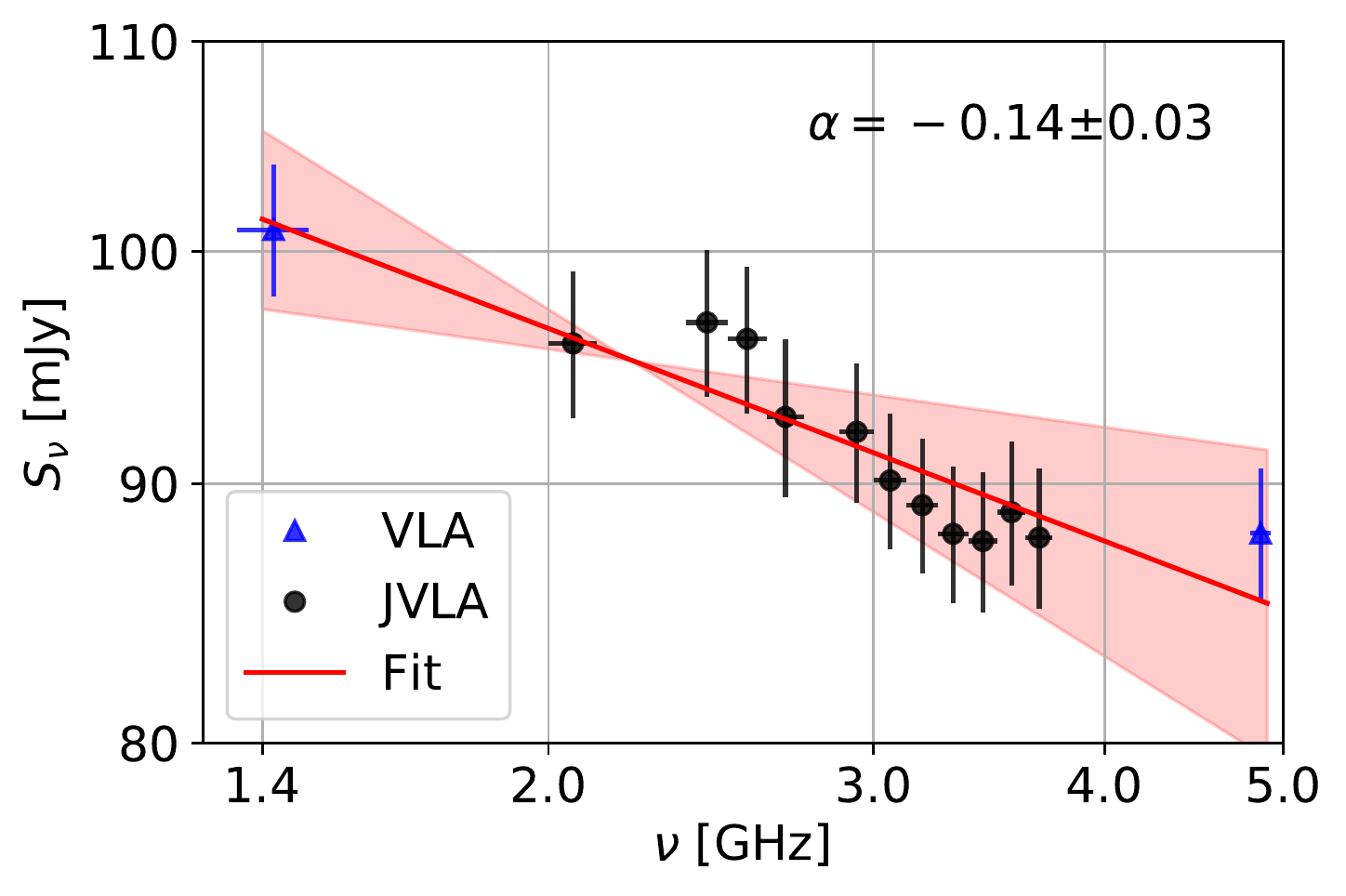}
    \caption{SED for the northern source ES. Filled lines represent 3-$\sigma$ uncertainties in the fit.}
    \label{fig:SED_NS}
\end{figure}

The same procedure was followed for the source ES. In this case we included in the fit the fluxes from \citet{Benaglia2010}, as this source is rather compact and we do not expect the emission from an H{\sc ii} region to vary. The value obtained for the spectral index is $\alpha=-0.14\pm0.03$ (Fig.\,\ref{fig:SED_NS}), which is consistent with the canonical index for ionised material in an optically thin regime ($\alpha = -0.1$) and with the value obtained by \cite{Benaglia2010}.

\begin{figure}
    \centering
    \includegraphics[width=0.99\linewidth]{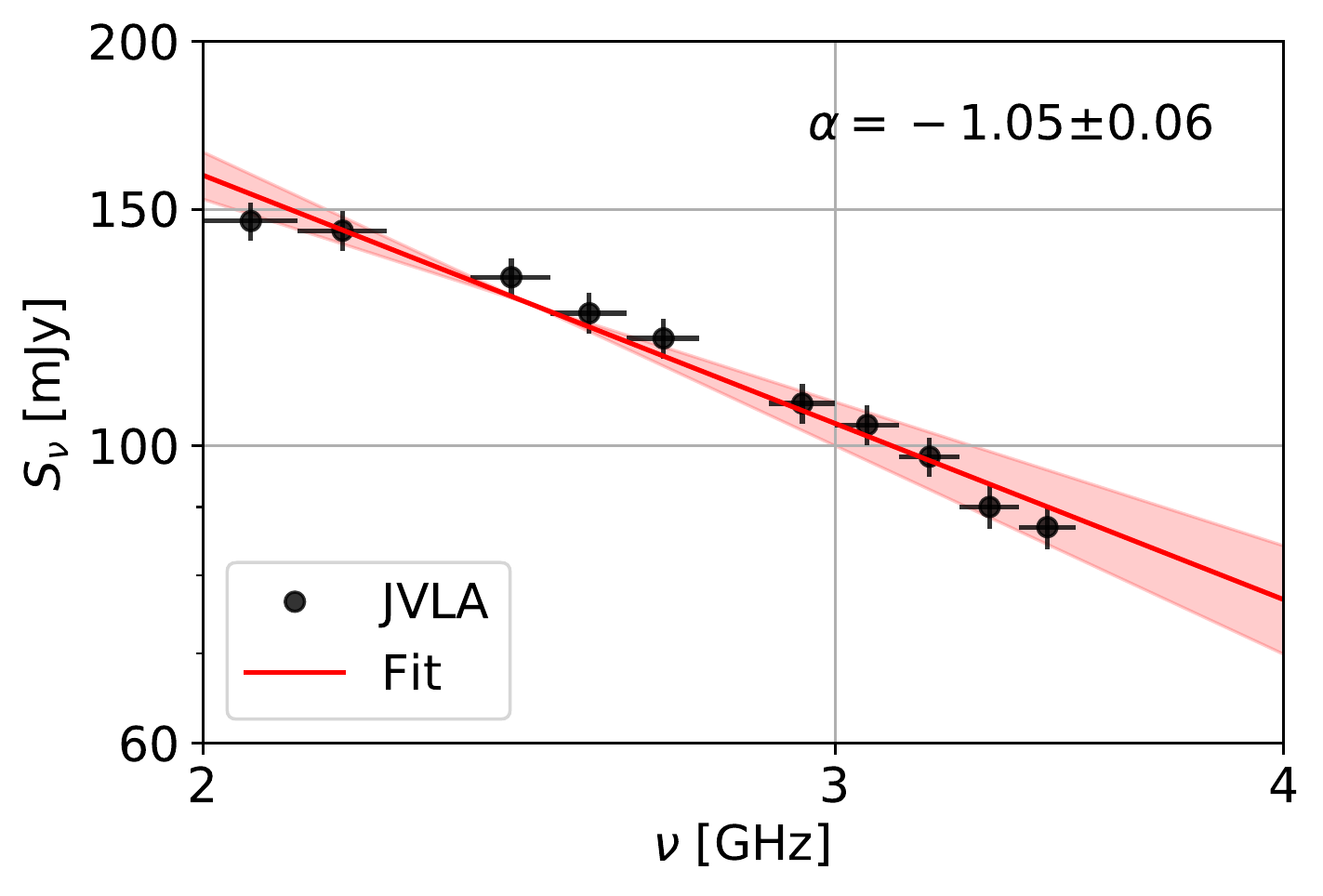}
    \caption{SED for the source here named S1. Filled lines represent 3-$\sigma$ uncertainties in the fit.}
    \label{fig:SED_SE}
\end{figure}

\begin{figure}
    \centering
    \includegraphics[width=0.99\linewidth]{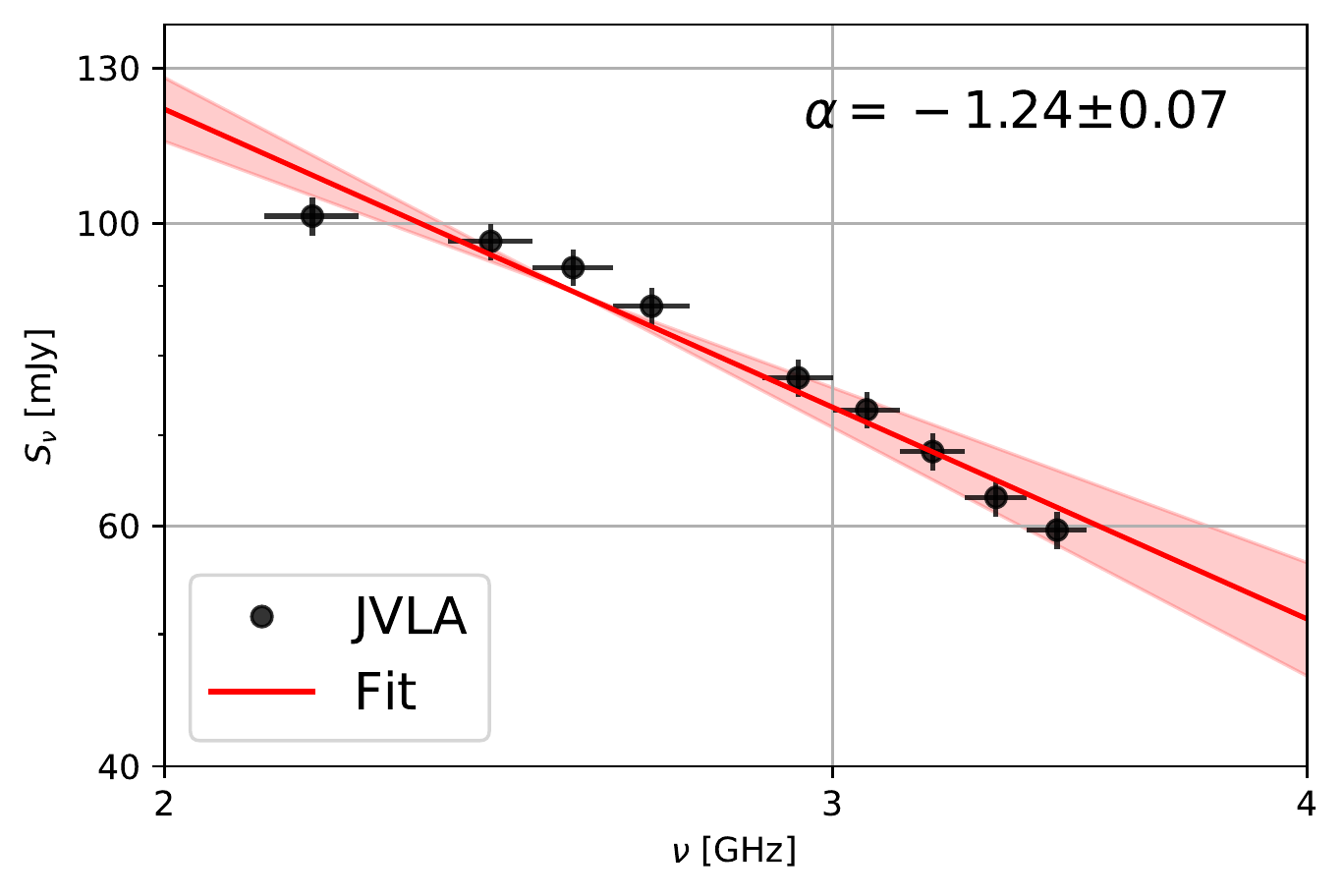}
    \caption{SED for the source here named S2. Filled lines represent 3-$\sigma$ uncertainties in the fit.}
    \label{fig:SED_SW}
\end{figure}

For the spectral fit for sources S1 and S2 we considered only the flux densities measured at spectral windows of S band, to exclude flux variability issues with non-coeval observations at other wavelengths. The SEDs and  their fits are shown in Figs.\,\ref{fig:SED_SE} and \ref{fig:SED_SW}. The resulting spectral indices are characteristic of extragalactic (non-thermal) sources: $\alpha_{\rm S1}=-1.05\pm0.06$, and $\alpha_{\rm S2}=-1.24\pm0.07$.

%------------------------------------------------
\subsection{RM synthesis and polarisation information}\label{sec:polinf}

\begin{figure}
\centering
	\includegraphics[angle=-0, width=0.95\linewidth]{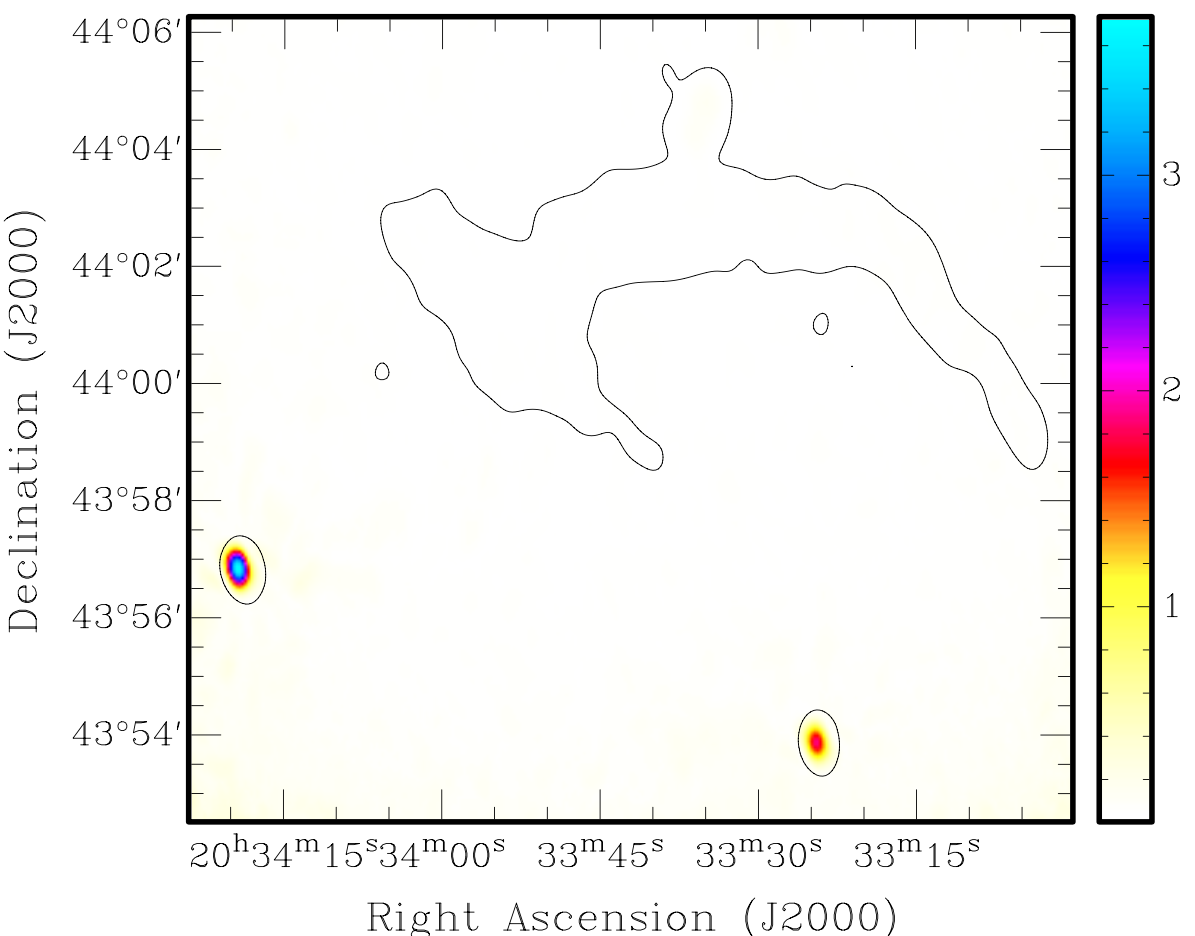}
    \caption{Polarised surface brightness modulus distribution (color scale, in mJy~beam$^{-1}$~RMTF$^{-1}$). In black, the 4-mJy~beam$^{-1}$ contour of the Stokes-I image convolved to a $34.5'' \times 21.0''$ beam.}
    \label{fig:polsurfbrigh}
\end{figure}

% Check of 3C286 pol yielded P/I = 10.3%.
% Its Phi -> 0 deg (we set +33deg at calibration)
Figure\,\ref{fig:polsurfbrigh} presents the polarised surface brightness modulus distribution \citep{Brentjens2005} resulting from RMS. The attained rms is 5\uJybeam. 

The Faraday depth can be expressed as the path integral $\phi(l)= 0.81\,\int{n_e \,B_{||} \,dl }$~rad\,m$^{-2}$, with $n_e$ the electron density in cm$^{-3}$, $B_{||}$ the magnetic field component parallel to the line-of-sight in $\mu$G, and $dl$ the differential path length in pc \citep{Brentjens2005}. For a uniform slab of size $\Delta l$ (in this case, the shocked stellar wind in the BS), we can approximate $|\phi| \approx \ 0.81\, n_e |B_{||}| \Delta l$, where $|B_{||}| \sim B/\sqrt{3}$ and $B$ is the magnetic field intensity \citep{Hales2017}.
A simplified estimate of $|\phi|$ from EB27, using the values of $n_e < 10$~cm$^{-3}$, $B < 140~\mu$G \citep{Benaglia2010}, and shock width $\Delta l \approx 0.3\,R_0 \approx 0.6$~pc \citep[][$R_0$ the stagnation point]{delValle2012}, yields $|\phi| < 500$~rad\,m$^{-2}$ (further discussion on the system parameters is provided in Sect.~\ref{sec:model}). This value serves as a reference for the expected Faraday depth of this source.

\begin{figure}
\centering
	\includegraphics[angle=-0, width=1.0\linewidth]{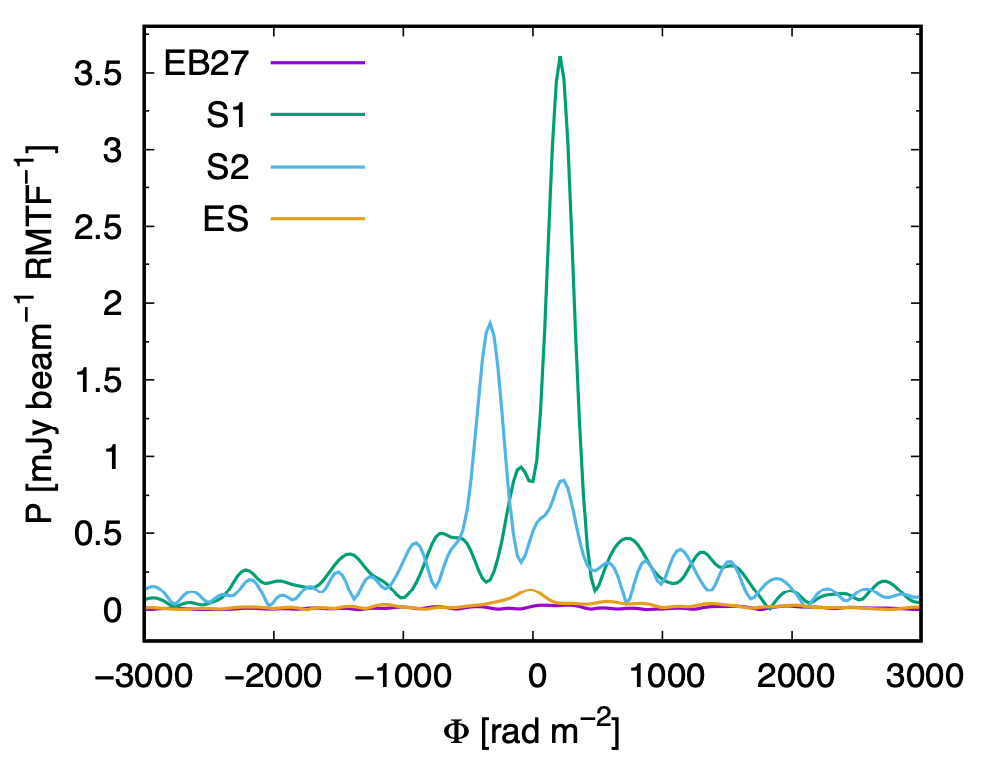}
	\caption{Cleaned Faraday depth profiles
	at four locations of the field-of-view: the core of the bow shock (EB27), and the sources ES, S1 and S2.} 
	% it is PB corrected.
\label{fig:phiprofiles}
\end{figure}

Along the Faraday depth interval $(-3000,3000)$~rad\,m$^{-2}$, $P$ was clearly detected for sources S1 and S2, and weakly for ES source. 
In Fig.\,\ref{fig:phiprofiles} we show the $\phi$ profiles for S1, S2, ES, and the heart of the BS. 
This last spectrum shows no peaks, and the values remain below 0.1~mJy~beam$^{-1}$~RMTF$^{-1}$. 
Fits to the maxima of the other three sources resulted in
$P_{\rm S1} =  3.6$~mJy~beam$^{-1}$~RMTF$^{-1}$ at $\phi=+210\pm13$~rad~m$^{-2}$; 
$P_{\rm S2} =  1.9$~mJy~beam$^{-1}$~RMTF$^{-1}$ at $\phi=-333\pm14$~rad~m$^{-2}$; 
$P_{\rm ES} =  0.1$~mJy~beam$^{-1}$~RMTF$^{-1}$ at $\phi=-25\pm18$~rad~m$^{-2}$. 
Taking into account the Stokes-I emission corresponding to the three sources, we get $D_{\rm S1} = 3.5\%$, $D_{\rm S2} = 2.4\%$, and $D_{\rm ES} = 0.35$\%.

With respect to the ES source, we note that it was formerly identified with an H{\sc ii} region (Sect.\,\ref{sec:intensity}). This is further supported by the spectral index we obtained ($\alpha = -0.14\pm0.03$; Fig.\,\ref{fig:SED_SE}), which is consistent with thermal emission from an optically thin plasma. In addition, its main and only $P$ feature is distributed around $\phi \sim 0$. All these facts lead us to interpret the bump in the ES spectrum as leakage. In this scenario, we can set $D_{\rm ES}$ as the fractional leakage, and use this value to characterise the error in $|P|$.
  
%======================================================
%
\section{Discussion}\label{sec:disc}
%
%======================================================

%-----------------------------------------------
\subsection{Emission model}\label{sec:model}
%-----------------------------------------------

The dominant radiative process at low radio frequencies is synchrotron radiation produced by relativistic electrons. Thus, the radio fluxes derived are valuable for constraining parameters related to the synchrotron luminosity of the BS. We highlight that the pioneer results by \cite{Benaglia2010} reported a flux of $\sim 100$~mJy for EB27 at 1.4~GHz, whereas in this study we obtain a flux of $\sim 700$~mJy at 2~GHz. Moreover, the corrected distance we presented here is (1.72/1.32) times larger than previously assumed. Overall, the intrinsic luminosity of the BS should be more than ten times larger than considered in earlier works.

We applied the emission model presented in \cite{delPalacio2018} in order to extract physical information from the radio SED of the BS. This model assumes that relativistic particles are accelerated at the reverse shock of the BS (the one that propagates in the stellar wind). At each position along the BS, the injected particle distribution is $Q(E)=Q_0\,E^{-p}\,\exp{(-E/E_\mathrm{max})}$, where $Q_0$ is a normalisation constant, $p$ is the spectral index (related to the radio spectral index as $p=-2\alpha+1$), and $E_\mathrm{max}$ is the maximum energy achieved by the particles. These particles are convected with the shocked flow along the BS 
while they suffer radiative and non-radiative losses. The thermodynamical properties of the gas along the BS are calculated using the Rankine-Hugoniot jump conditions for a strong shock. In addition, a prescription for the magnetic field is adopted such that the magnetic field pressure is a fraction $\zeta_B$ of the thermal pressure of the gas. At each position along the BS, the synchrotron luminosity is calculated using the computed particle population and the magnetic field intensity. This allows us to generate 3-D emission maps that are projected and convolved with a gaussian beam in order to produce synthetic emission maps. For further details on the model we refer to \cite{delPalacio2018}.
  
Regarding the model parameters, we need a set that can reconcile the high value of $R_\mathrm{0,proj} \approx 3.9\arcmin$ ($R_0 \approx 1.9$~pc), the non-detection of the free-free emission of the stellar wind, and the high synchrotron luminosity of the BS ($L_\mathrm{sy} \approx 6\times10^{33}$~erg\,s$^{-1}$). The value of $R_0$ is given by the pressure balance between the stellar wind and the ambient ram pressure. The most relevant parameters involved are: the density of the medium ($n_\mathrm{ISM}$), the stellar wind velocity ($v_\infty$) and mass-loss rate ($\dot{M}$), the peculiar velocity of the star ($V_\star$), the source distance ($d$), the fraction of the stellar wind power injected into non-thermal electrons ($f_\mathrm{NT,e}$), and the intensity of the magnetic field in the BS ($B$); we refer to \cite{delPalacio2018} for the scaling between these parameters and $R_0$ and $L_\mathrm{sy}$. 

According to the velocity calculated in Sect~\ref{sec:EB27}, we have $V_\star > 44$~km\,s$^{-1}$. We adopt a rather high value of $v_\infty = 3000$~km\,s$^{-1}$, which can be expected in this type of stars. \textcolor{red}{\cite{Kobulnicky2018}} estimated that the ambient density was $n_\mathrm{ISM} \approx 19$~cm$^{-3}$, but it should be decreased at least by a factor 1.32/1.72 (due to the correction in the distance to the star), which yields $n_\mathrm{ISM}=15$~cm$^{-3}$. However, this density is too high as it needs to be compensated by an excessive value of $\dot{M} \approx 2\times10^{-5}$\Msunyr, which does not satisfy the radio flux constraints (Sec.~\ref{sec:intensity}). We therefore adopt $n_\mathrm{ISM}=9$~cm$^{-3}$ and $\dot{M} \approx 10^{-5}$\Msunyr as consistent values that can reconcile the observational constraints. 

From the spectral index maps (Fig.~\ref{fig:alpha-y-error}), in the apex of the BS the emission has an index $\alpha \approx -0.6$, which translates in an injection index of $p \approx 2.2$ (softer than the value $p=2$ previously considered). The regions with a more negative spectral index can be caused by: i) efficient cooling of the electrons, ii) efficient diffusion, iii) an observational bias towards underestimating the flux at higher frequencies. In principle, convective escape is expected to dominate the losses for electrons \citep{delPalacio2018}, though diffusion escape could also be relevant depending on the unknown diffusion regime \citep{DelValle2018}. In addition, the softer electron energy distribution implies a smaller high-energy flux, which is easier to reconcile with the non-detection in X-rays by \cite{Toala2016} and in $\gamma$-rays by \cite{Schulz2014}.

%-----------------------------------------------
\subsection{Synthetic emission maps}
%-----------------------------------------------

\begin{figure*}
	\includegraphics[width=0.5\linewidth]{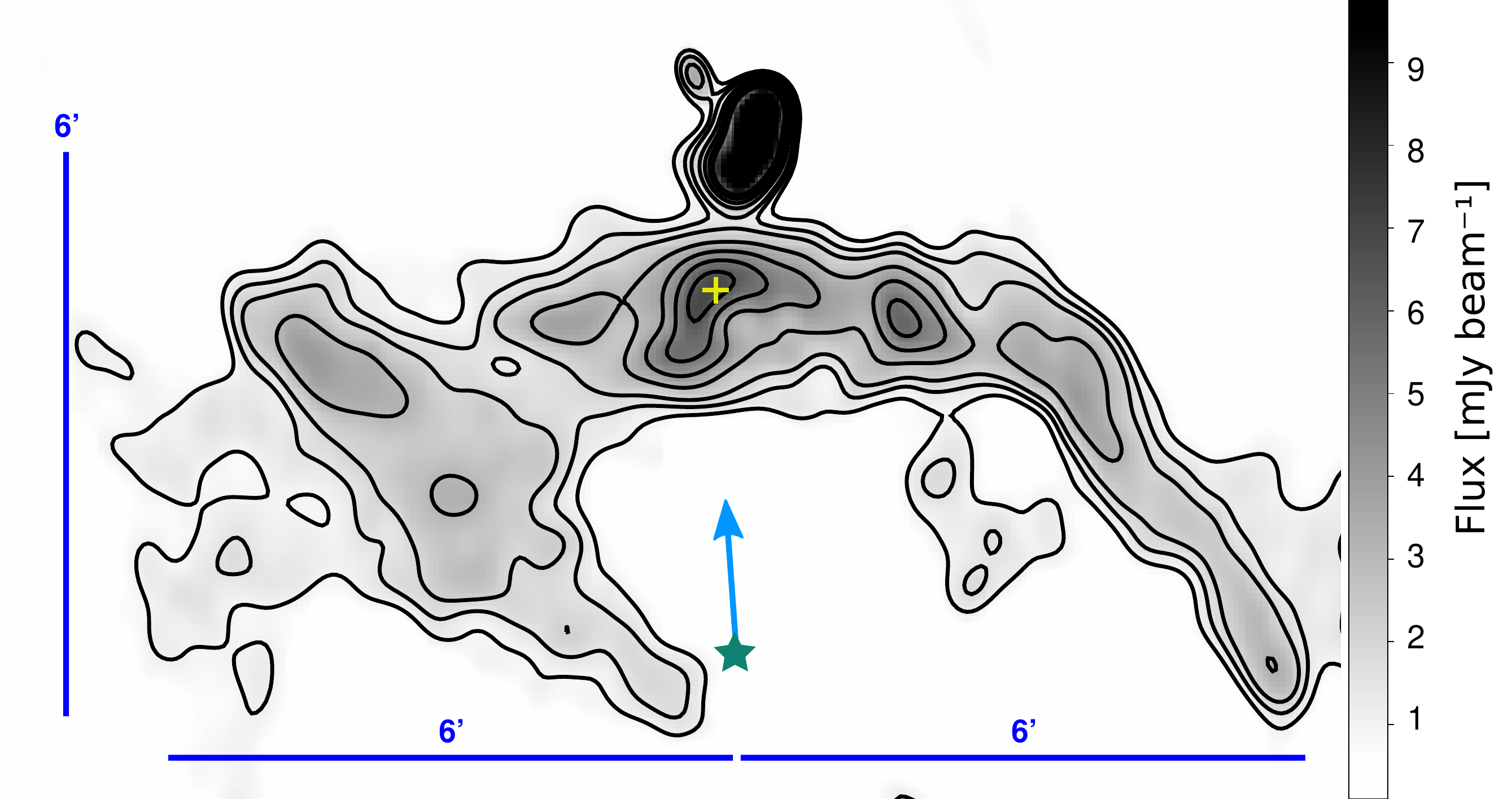} \\
	\includegraphics[angle=270, width=0.49\linewidth]{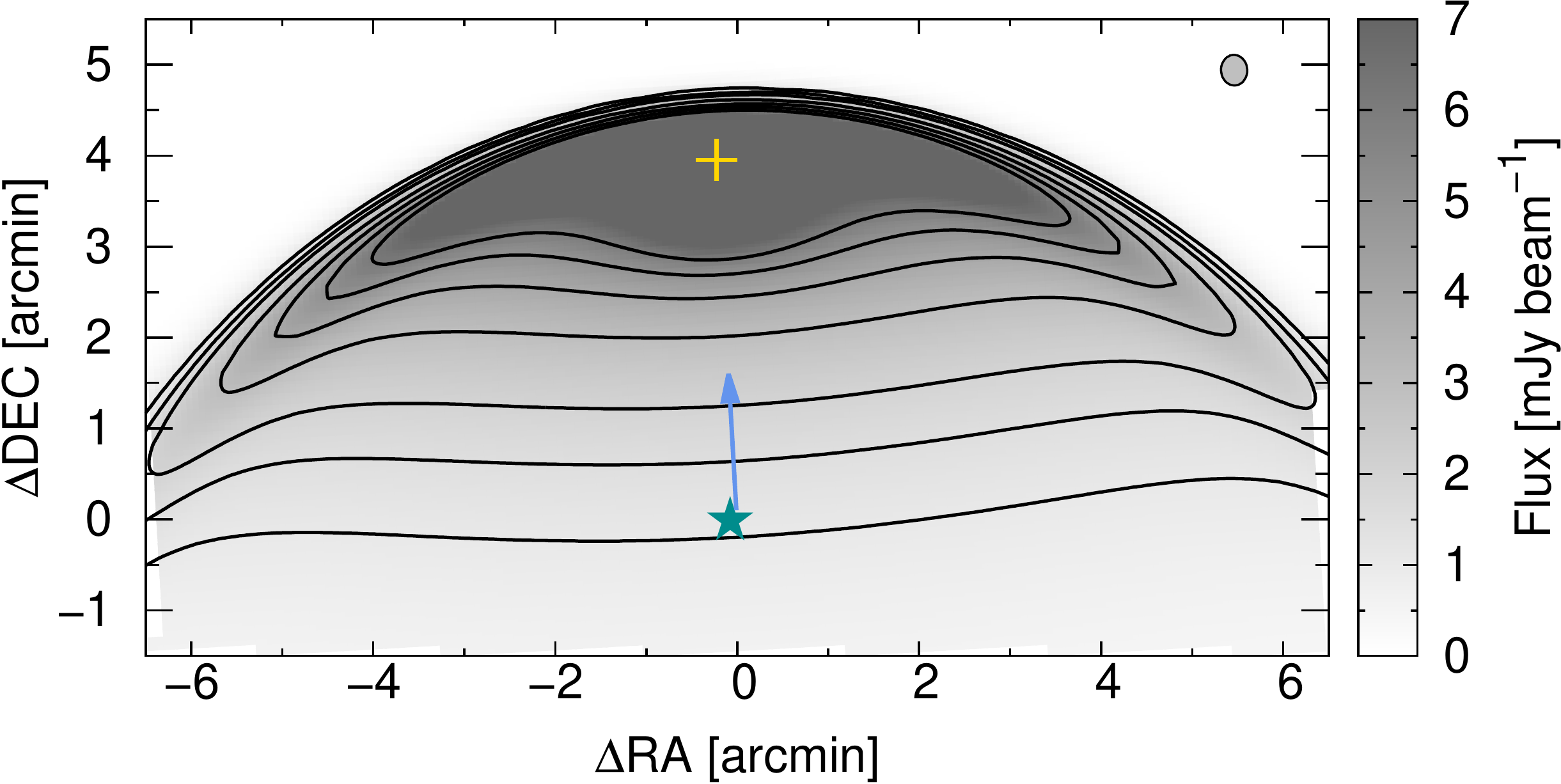} 
	\includegraphics[angle=270, width=0.49\linewidth]{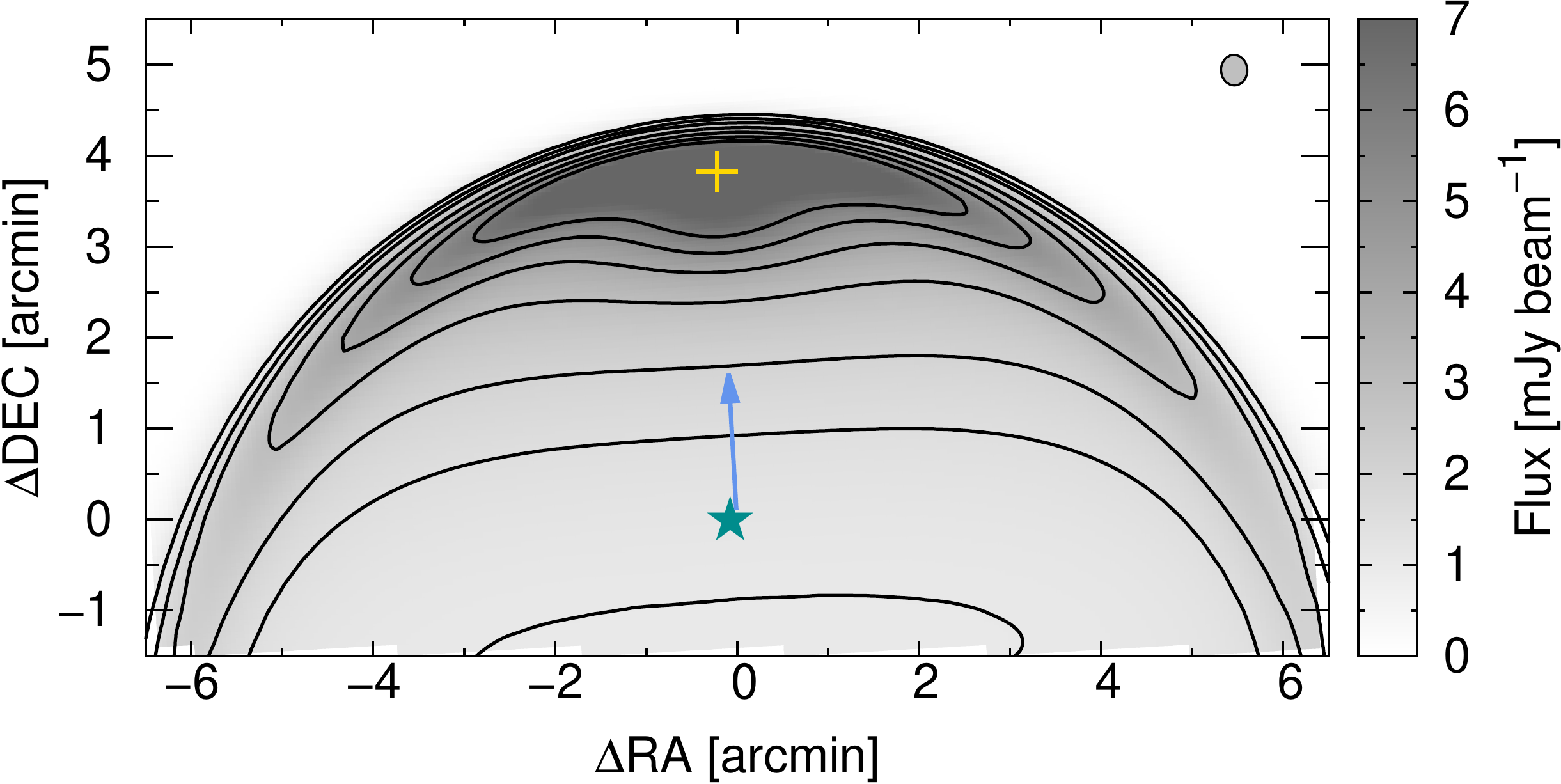}
    \caption{Observed map (top) and simulated maps (bottom). Bottom left is for $r=0.01$ and bottom right for $r=0.3$. Contour levels are 1, 1.5, 2, 3, 4, 5, and 6\mJybeam. }
    \label{fig:synthetic_maps}
\end{figure*}

The observed morphology of the BS provides further information that can be taken into account in an extended emission model. We therefore produce synthetic emission maps by convolving the modeled emission from the BS with a gaussian beam of $20.2''\times12.5''$. These maps are compared with the observed maps in Fig.~\ref{fig:synthetic_maps}.

One of the most relevant parameters in shaping the observed morphology of the BS is the inclination angle between the line of sight and the peculiar velocity of the star. This angle can be estimated as $i \approx \arctan{(V_r/V_t)}$, where $V_t$ and $V_r$ are the tangential and radial components of the stellar velocity with respect to its surrounding medium. We derived a value of $V_t \approx 43.6$\kms (Sect.~\ref{sec:EB27}), but the value of $V_r$ is poorly constrained. \cite{Kobulnicky2010} reported a radial velocity of $\sim -66$\kms, but this value has a large uncertainty and it is not corrected for the motion of the surrounding medium. Here we show that the observed morphology is consistent with an angle $i \sim 60\degr$, which translates in $V_r \sim -25$\kms and $V_\star \approx 50$\kms.

The shape of the BS is also determined by the hydrodynamics of the interaction between the stellar wind, the ISM, and the shocked material. The model by \cite{delPalacio2018} uses the analytical prescriptions for the geometry of the BS given in \cite{Christie2016}. In this formalism, the ISM can have a non-negligible pressure \citep[unlike the solution by][that is valid only for a cold ISM]{Wilkin1996} that helps to collimate the BS structure. This is incorporated through a factor $r$ defined as the ratio between the thermal pressure in the ISM and the ram pressure in the ISM ($\mathcal{P}_\mathrm{kin} \propto n_\mathrm{ISM} V_\star^2$). The value of $r$ is related to the Mach number as $M_\mathrm{s} = (\gamma_\mathrm{ad} r)^{-1/2}$, with $\gamma_\mathrm{ad}=5/3$ the adiabatic coefficient. A value of $r \ll 1$ corresponds to a cold medium that exerts a negligible pressure (or to $M_\mathrm{s} \gg 1$), whereas for a warm, ionised ISM with temperature 10\,000~K we expect $r \approx 0.07$ (or $M_\mathrm{s} \approx 2.9$).
In Fig.~\ref{fig:synthetic_maps} we show that the observed morphology is more consistent with a warm medium exerting a significant pressure such that $r \sim 0.3$. Nonetheless, other effects such as a high stellar magnetic field can also have an influence in the shape of the BS \citep{Meyer2017, Mackey2020}.

%-----------------------------------------------
\subsection{Magnetic field intensity}
%-----------------------------------------------

To fit the observed flux of $\approx 500$~mJy at 3~GHz, we evaluate two extreme cases. First, we consider a high magnetic field value such that the magnetic field pressure is equal to the thermal pressure in the shocked wind ($\zeta_B=1$). Higher values of the magnetic field would render the fluid incompressible and therefore no shock would form, which in turns means no diffusive shock acceleration would be possible. This equal pressure condition yields $B \approx 100~\mu$G. To match the observed flux, it is required that a fraction $f_\mathrm{NT,e} \approx 1.5\%$ of the available stellar wind kinetic power is transferred to relativistic electrons. Assuming that the kinetic power injected in relativistic protons is ten times larger than in electrons, the ratio between the energy density in relativistic particles and the energy density in the magnetic field ranges from $U_\mathrm{NT}/U_\mathrm{B} \sim 10^{-5}$ to $10^{-3}$ along the bow shock. In this case, the high $B$ values should be attained by magnetohydrodynamical processes \citep[e.g., adiabatic compression of stellar magnetic field lines,][]{delPalacio2018} and not by amplification by the cosmic rays. This requirement can be met by a surface stellar magnetic field of $B_\star \approx 2.5 B \, (R_0/R_\star) \approx 360$~G.

Second, we consider an extreme case in which the injected energy in relativistic electrons is $f_\mathrm{NT,e} \approx 10\%$, from which we obtain $B \approx 35~\mu$G to match the observed fluxes. Lower values of the magnetic field would require an excessively efficient conversion of wind kinetic power into relativistic particles. In this case, we get $U_\mathrm{NT}/U_\mathrm{B} \sim 5\times10^{-4}$ to $5\times10^{-2}$ along the bow shock. Once again, this $B$ values are not likely to be achieved by magnetic field amplification by the cosmic rays. Instead, a high stellar magnetic field $B_\star \approx 1$~kG would be required.

We thus conclude that the magnetic field in the BS should be $B \sim 35$--100~$\mu$G, and that electrons are efficiently accelerated obtaining 1--10\% of the available wind kinetic power in the BS. The non-thermal particle population is in energy subequipartition with respect to the magnetic field ($U_\mathrm{NT} \ll U_\mathrm{B}$).

\subsection{Non-detection of polarisation in EB27}

In Sect.~\ref{sec:polinf} we obtained information of the observed polarisation of EB27 and other sources in the field. Given a relativistic electron population of the form $N(E) \propto E^{-p}$, the intrinsic linear polarisation of their synchrotron spectrum is $D_{\rm i}(p) = (3p+3)/(3p+7)$. Notwithstanding, a synchrotron source can have a lower polarisation due to a series of factors. Following \citet{Hales2017} and references therein, a turbulent magnetic field greatly decreases the observed polarisation degree $D_{\rm obs}$:
% valid in the case of an unresolved source 
%
\begin{equation}
%\nonumber
 D_\mathrm{obs}(p, \nu)= D_\mathrm{i}(p) \, \frac{B_0^2}{B_0^2+B_\mathrm{r}^2} \, \xi(\nu),
\end{equation}
where $B_\mathrm{r}$ and $B_0$ are the random and ordered components of the magnetic field, respectively, and the function $\xi(\nu)$ takes into account additional depolarisation effects. These effects are either frequency-independent or stronger at shorter frequencies. Most depolarisation mechanisms have significant effects in the regime where $|\phi|\,\lambda^2 > 1$, so that the value of $|\phi|\,\lambda^2$ helps to assess whether depolarisation is likely to be important. For sources S1 and S2 $\phi$-values were derived, but for EB27, since its Faraday depth profile resulted in noise, no information on $\phi$ could be retrieved.

If depolarisation effects are neglected, the random and ordered components of the field relate with the observed polarisation as $B_\mathrm{r}/B_0 = \sqrt{ D_{\rm i}/D_\mathrm{obs} - 1}$. Applying this formula to EB27, for which $D_{\rm i} \gtrsim 70\%$ and $D_\mathrm{obs} < 0.5\%$, we get $B_r/B_0 > 12$. Thus, the very low value of $D_{\rm obs}$ can be the footprint of a highly turbulent medium, in which the diffusion shock acceleration mechanism is unlikely and other mechanisms such as turbulent magnetic reconnection could be at work \citep[][and references therein]{Hales2017}.

Previous works on BSs associated with pulsars showed that their magnetic fields are usually well-ordered, leading to high polarisation degrees \citep[see for instance][]{Kulkarni1992,Reich1992,Yusef-Zadeh2005,Ng2012}. However, pulsars have stronger magnetic fields than stars, which can significantly reduce the turbulence in the BS \citep[e.g.][for the case of stellar BSs]{vanmarle2014}. In addition, pulsar winds are much lighter than the winds from massive stars, which can also lead to a less significant Faraday rotation and consequent depolarisation. Thus, it is reasonable to expect a lower polarisation degree in a BS associated to a massive star.

In the case of the sources S1 and S2, we get $p=3.1$, $D_{\rm i} = 75.5\%$, and  $p=3.48$, $D_{\rm i} = 77.1\%$ respectively; these values are characteristic of extragalactic (active galactic nuclei) sources. However, for these sources, neglecting depolarisation effects is likely to be misleading due to their high $|\phi|$ values.

Among the possible depolarisation effects, one can list the contribution from thermal (un-polarised) emission, differential (depth) Faraday rotation, internal Faraday dispersion, external Faraday dispersion, and (for unresolved sources) gradients in RM across the beam; see the detailed description given in Sect.~4.2 of \cite{Hales2017} in the context of shocks in massive colliding-wind binaries. Nonetheless, none of these seem to be relevant in the BS scenario (studied here) given that the source is resolved and the plasma is more diluted and with lower magnetic fields than in the scenario analysed by \cite{Hales2017}. On the contrary, some of these depolarisation mechanisms are likely to have a significant effect in the observed polarisation for sources S1 and S2, since for both sources $|\phi| \lambda^2 > 1$ (for S1, $|\phi_1| \lambda^2 \approx 2.1$, and for S2, $|\phi_2| \lambda^2 \approx 3.3$).

%======================================================
%
\section{Conclusions}\label{sec:conclusions}
%
%======================================================
%
%This section should briefly summarise what has been done, and describe the final conclusions which the authors draw from their work.

Wide S-band radio observations towards the object EB27 allowed us to better characterise this source and its surroundings. 

By taking advantage of the latest results from the \textit{Gaia} mission, we found that the corrected proper motion now indicates that the apex of the bow shock from \BD coincides with its brightest spot, as expected. Moreover, the trajectory of the star coincides with it having been ejected from the centre of Cyg~OB2 roughly 1.5~Myr ago.

The upper limit of the flux density measured at the stellar position discards relevant non-thermal emission from a putative colliding wind region, and thus that BD+43$^\circ$3654 is either a runaway single star or perhaps a very tight system. This upper limit also sets constraints to the stellar mass-loss rate.

We  were able to detach EB27 from the surrounding sources, and helped also by previous studies and measurements, to confirm they are at very different distances and thus physically unrelated. We gathered additional evidence towards the extragalactic origin of sources here named S1 and S2, and  
derived the observed rotation angle at S-band for both of them. We corroborated the nature of the H{\sc ii} region right north, though farther away from EB27.

Our deeper radio observation revealed that the BS is steeper and brighter than previously reported. On the one hand, together with the larger distance than previously assumed, this requires an intrinsically brighter emission. This, in turn, suggests that the stellar wind is more powerful and massive that usually assumed. On the other hand, the steeper radio spectrum implies a smaller high-energy flux, which is easier to reconcile with the non-detection in X-rays and $\gamma$-rays.

Finally, the polarimetric study with unprecendented sensitivity in the EB27 kind of objects showed a degree of linear polarisation below 1\%; the steps followed to derive these results are validated by polarisation measurements of S1 and S2 at a $\sim$10 signal-to-noise level. A highly turbulent magnetic field and/or Faraday rotation in a diffuse, ill-imaged medium on this galactic plane region full of emission at different angular scales are possible causes of the absence of detection. We tentatively suggest that non-polarised synchrotron emission might be a common denominator to systems involving massive stars.

\section*{Acknowledgements}

%The Acknowledgements section is not numbered. Here you can thank helpful colleagues, acknowledge funding agencies, telescopes and facilities used etc. Try to keep it short.

PB acknowledges support from ANPCyT PICT-2017 0773 and from the staff of the Jansky Very Large Array during her stay.
This work has made use of the Simbad database (operated at CDS, Strasbourg, France), of NASA's Astrophysics Data System bibliographic services, of Astropy \citep[a community-developed core Python package for Astronomy, http://www.astropy.org,][]{astropy2013, astropy2018},  and of data from the European Space Agency (ESA) mission {\it Gaia} (\url{https://www.cosmos.esa.int/gaia}), processed by the {\it Gaia} Data Processing and Analysis Consortium (DPAC,
\url{https://www.cosmos.esa.int/web/gaia/dpac/consortium}); funding for the DPAC has been provided by national institutions, in particular the institutions participating in the {\it Gaia} Multilateral Agreement.  

%%%%%%%%%%%%%%%%%%%%%%%%%%%%%%%%%%%%%%%%%%%%%%%%%%
\section*{Data Availability}

The radio data studied here can be retrieved at the URL https://archive.nrao.edu/archive/advquery.jsp, under the project code 18A-168.
%The inclusion of a Data Availability Statement is a requirement for articles published in MNRAS. Data Availability Statements provide a standardised format for readers to understand the availability of data underlying the research results described in the article. The statement may refer to original data generated in the course of the study or to third-party data analysed in the article. The statement should describe and provide means of access, where possible, by linking to the data or providing the required accession numbers for the relevant databases or DOIs.

%%%%%%%%%%%%%%%%%%%% REFERENCES %%%%%%%%%%%%%%%%%%

\bibliographystyle{mnras}
\bibliography{references} 

\begin{thebibliography}{}
\makeatletter
\relax
\def\mn@urlcharsother{\let\do\@makeother \do\$\do\&\do\#\do\^\do\_\do\%\do\~}
\def\mn@doi{\begingroup\mn@urlcharsother \@ifnextchar [ {\mn@doi@}
  {\mn@doi@[]}}
\def\mn@doi@[#1]#2{\def\@tempa{#1}\ifx\@tempa\@empty \href
  {http://dx.doi.org/#2} {doi:#2}\else \href {http://dx.doi.org/#2} {#1}\fi
  \endgroup}
\def\mn@eprint#1#2{\mn@eprint@#1:#2::\@nil}
\def\mn@eprint@arXiv#1{\href {http://arxiv.org/abs/#1} {{\tt arXiv:#1}}}
\def\mn@eprint@dblp#1{\href {http://dblp.uni-trier.de/rec/bibtex/#1.xml}
  {dblp:#1}}
\def\mn@eprint@#1:#2:#3:#4\@nil{\def\@tempa {#1}\def\@tempb {#2}\def\@tempc
  {#3}\ifx \@tempc \@empty \let \@tempc \@tempb \let \@tempb \@tempa \fi \ifx
  \@tempb \@empty \def\@tempb {arXiv}\fi \@ifundefined
  {mn@eprint@\@tempb}{\@tempb:\@tempc}{\expandafter \expandafter \csname
  mn@eprint@\@tempb\endcsname \expandafter{\@tempc}}}

\bibitem[\protect\citeauthoryear{{Astropy Collaboration} et~al.,}{{Astropy
  Collaboration} et~al.}{2013}]{astropy2013}
{Astropy Collaboration} et~al., 2013, \mn@doi [\aap]
  {10.1051/0004-6361/201322068}, \href
  {https://ui.adsabs.harvard.edu/abs/2013A&A...558A..33A} {558, A33}

\bibitem[\protect\citeauthoryear{{Astropy Collaboration} et~al.,}{{Astropy
  Collaboration} et~al.}{2018}]{astropy2018}
{Astropy Collaboration} et~al., 2018, \mn@doi [\aj] {10.3847/1538-3881/aabc4f},
  \href {https://ui.adsabs.harvard.edu/abs/2018AJ....156..123A} {156, 123}

\bibitem[\protect\citeauthoryear{{Bell}}{{Bell}}{1978}]{Bell1978}
{Bell} A.~R.,  1978, \mn@doi [\mnras] {10.1093/mnras/182.2.147}, \href
  {https://ui.adsabs.harvard.edu/abs/1978MNRAS.182..147B} {182, 147}

\bibitem[\protect\citeauthoryear{{Benaglia}, {Romero}, {Mart{\'\i}}, {Peri}  \&
  {Araudo}}{{Benaglia} et~al.}{2010}]{Benaglia2010}
{Benaglia} P.,  {Romero} G.~E.,  {Mart{\'\i}} J.,  {Peri} C.~S.,   {Araudo}
  A.~T.,  2010, \mn@doi [\aap] {10.1051/0004-6361/201015232}, \href
  {https://ui.adsabs.harvard.edu/abs/2010A&A...517L..10B} {517, L10}

\bibitem[\protect\citeauthoryear{{Benaglia}, {De Becker}, {Ishwara-Chandra},
  {Intema}  \& {Isequilla}}{{Benaglia} et~al.}{2020}]{Benaglia2020a}
{Benaglia} P.,  {De Becker} M.,  {Ishwara-Chandra} C.~H.,  {Intema} H.~T.,
  {Isequilla} N.~L.,  2020, \mn@doi [\pasa] {10.1017/pasa.2020.21}, \href
  {https://ui.adsabs.harvard.edu/abs/2020PASA...37...30B} {37, e030}

\bibitem[\protect\citeauthoryear{{Berlanas}, {Wright}, {Herrero}, {Drew}  \&
  {Lennon}}{{Berlanas} et~al.}{2019}]{Berlanas2019}
{Berlanas} S.~R.,  {Wright} N.~J.,  {Herrero} A.,  {Drew} J.~E.,   {Lennon}
  D.~J.,  2019, \mn@doi [\mnras] {10.1093/mnras/stz117}, \href
  {https://ui.adsabs.harvard.edu/abs/2019MNRAS.484.1838B} {484, 1838}

\bibitem[\protect\citeauthoryear{{Bobylev} \& {Bajkova}}{{Bobylev} \&
  {Bajkova}}{2019}]{Bobylev2019}
{Bobylev} V.~V.,  {Bajkova} A.~T.,  2019, \mn@doi [Astronomy Letters]
  {10.1134/S1063773719040029}, \href
  {https://ui.adsabs.harvard.edu/abs/2019AstL...45..208B} {45, 208}

\bibitem[\protect\citeauthoryear{{Brentjens} \& {de Bruyn}}{{Brentjens} \& {de
  Bruyn}}{2005}]{Brentjens2005}
{Brentjens} M.~A.,  {de Bruyn} A.~G.,  2005, \mn@doi [\aap]
  {10.1051/0004-6361:20052990}, \href
  {https://ui.adsabs.harvard.edu/abs/2005A&A...441.1217B} {441, 1217}

\bibitem[\protect\citeauthoryear{{Brookes}}{{Brookes}}{2016}]{Brookes2016}
{Brookes} D.~P.,  2016, PhD thesis, University of Birmingham,
  \mn@doi{10.5281/zenodo.60136}

\bibitem[\protect\citeauthoryear{{Cameron} \& {Kulkarni}}{{Cameron} \&
  {Kulkarni}}{2007}]{Cameron2007}
{Cameron} P.~B.,  {Kulkarni} S.~R.,  2007, \mn@doi [\apjl] {10.1086/521077},
  \href {https://ui.adsabs.harvard.edu/abs/2007ApJ...665L.135C} {665, L135}

\bibitem[\protect\citeauthoryear{{Christie}, {Petropoulou}, {Mimica}  \&
  {Giannios}}{{Christie} et~al.}{2016}]{Christie2016}
{Christie} I.~M.,  {Petropoulou} M.,  {Mimica} P.,   {Giannios} D.,  2016,
  \mn@doi [\mnras] {10.1093/mnras/stw749}, \href
  {http://adsabs.harvard.edu/abs/2016MNRAS.459.2420C} {459, 2420}

\bibitem[\protect\citeauthoryear{Collaboration, Brown, Vallenari, Prusti, de
  Bruijne, Babusiaux  \& Biermann}{Collaboration et~al.}{2020}]{Brown2020b}
Collaboration G.,  Brown A. G.~A.,  Vallenari A.,  Prusti T.,  de Bruijne J.
  H.~J.,  Babusiaux C.,   Biermann M.,  2020, Gaia Early Data Release 3:
  Summary of the contents and survey properties (\mn@eprint {arXiv}
  {2012.01533})

\bibitem[\protect\citeauthoryear{{Comer{\'o}n} \& {Pasquali}}{{Comer{\'o}n} \&
  {Pasquali}}{2007}]{Comeron2007}
{Comer{\'o}n} F.,  {Pasquali} A.,  2007, \mn@doi [\aap]
  {10.1051/0004-6361:20077304}, \href
  {http://adsabs.harvard.edu/abs/2007A%26A...467L..23C} {467, L23}

\bibitem[\protect\citeauthoryear{{Condon}, {Cotton}, {Greisen}, {Yin},
  {Perley}, {Taylor}  \& {Broderick}}{{Condon} et~al.}{1998}]{Condon1998}
{Condon} J.~J.,  {Cotton} W.~D.,  {Greisen} E.~W.,  {Yin} Q.~F.,  {Perley}
  R.~A.,  {Taylor} G.~B.,   {Broderick} J.~J.,  1998, \mn@doi [\aj]
  {10.1086/300337}, \href
  {https://ui.adsabs.harvard.edu/abs/1998AJ....115.1693C} {115, 1693}

\bibitem[\protect\citeauthoryear{{De Becker}, {del Valle}, {Romero}, {Peri}  \&
  {Benaglia}}{{De Becker} et~al.}{2017}]{DeBecker2017}
{De Becker} M.,  {del Valle} M.~V.,  {Romero} G.~E.,  {Peri} C.~S.,
  {Benaglia} P.,  2017, \mn@doi [\mnras] {10.1093/mnras/stx1826}, \href
  {https://ui.adsabs.harvard.edu/abs/2017MNRAS.471.4452D} {471, 4452}

\bibitem[\protect\citeauthoryear{{Douglas}, {Bash}, {Bozyan}, {Torrence}  \&
  {Wolfe}}{{Douglas} et~al.}{1996}]{Douglas1996}
{Douglas} J.~N.,  {Bash} F.~N.,  {Bozyan} F.~A.,  {Torrence} G.~W.,   {Wolfe}
  C.,  1996, \mn@doi [\aj] {10.1086/117932}, \href
  {https://ui.adsabs.harvard.edu/abs/1996AJ....111.1945D} {111, 1945}

\bibitem[\protect\citeauthoryear{{Dyson}}{{Dyson}}{1975}]{Dyson1975}
{Dyson} J.~E.,  1975, \mn@doi [\apss] {10.1007/BF00636999}, \href
  {http://adsabs.harvard.edu/abs/1975Ap%26SS..35..299D} {35, 299}

\bibitem[\protect\citeauthoryear{{Gaia Collaboration} et~al.,}{{Gaia
  Collaboration} et~al.}{2016}]{Gaia2016b}
{Gaia Collaboration} et~al., 2016, \mn@doi [\aap]
  {10.1051/0004-6361/201629272}, \href
  {https://ui.adsabs.harvard.edu/abs/2016A&A...595A...1G} {595, A1}

\bibitem[\protect\citeauthoryear{{H.~E.~S.~S. Collaboration}
  et~al.,}{{H.~E.~S.~S. Collaboration} et~al.}{2018}]{HESS2018}
{H.~E.~S.~S. Collaboration} et~al., 2018, \mn@doi [\aap]
  {10.1051/0004-6361/201630151}, \href
  {https://ui.adsabs.harvard.edu/abs/2018A&A...612A..12H} {612, A12}

\bibitem[\protect\citeauthoryear{{Hales} \& {Middelberg}}{{Hales} \&
  {Middelberg}}{2014}]{Hales2014}
{Hales} C.~A.,  {Middelberg} E.,  2014, {pieflag: CASA task to efficiently flag
  bad data} (\mn@eprint {ascl} {1408.014})

\bibitem[\protect\citeauthoryear{{Hales}, {Benaglia}, {del Palacio}, {Romero}
  \& {Koribalski}}{{Hales} et~al.}{2017}]{Hales2017}
{Hales} C.~A.,  {Benaglia} P.,  {del Palacio} S.,  {Romero} G.~E.,
  {Koribalski} B.~S.,  2017, \mn@doi [A\&A] {10.1051/0004-6361/201629644},
  \href {http://adsabs.harvard.edu/abs/2017A%26A...598A..42H} {598, A42}

\bibitem[\protect\citeauthoryear{{Heald}}{{Heald}}{2009}]{heald2009}
{Heald} G.,  2009, in {Strassmeier} K.~G.,  {Kosovichev} A.~G.,   {Beckman}
  J.~E.,  eds,  IAU Symposium Vol. 259, Cosmic Magnetic Fields: From Planets,
  to Stars and Galaxies. pp 591--602, \mn@doi{10.1017/S1743921309031421}

\bibitem[\protect\citeauthoryear{{Kobulnicky}, {Gilbert}  \&
  {Kiminki}}{{Kobulnicky} et~al.}{2010}]{Kobulnicky2010}
{Kobulnicky} H.~A.,  {Gilbert} I.~J.,   {Kiminki} D.~C.,  2010, \mn@doi [\apj]
  {10.1088/0004-637X/710/1/549}, \href
  {https://ui.adsabs.harvard.edu/abs/2010ApJ...710..549K} {710, 549}

\bibitem[\protect\citeauthoryear{{Kobulnicky} et~al.,}{{Kobulnicky}
  et~al.}{2016}]{Kobulnicky2016}
{Kobulnicky} H.~A.,  et~al., 2016, \mn@doi [\apjs]
  {10.3847/0067-0049/227/2/18}, \href
  {https://ui.adsabs.harvard.edu/abs/2016ApJS..227...18K} {227, 18}

\bibitem[\protect\citeauthoryear{{Kobulnicky}, {Chick}  \&
  {Povich}}{{Kobulnicky} et~al.}{2018}]{Kobulnicky2018}
{Kobulnicky} H.~A.,  {Chick} W.~T.,   {Povich} M.~S.,  2018, \mn@doi [\apj]
  {10.3847/1538-4357/aab3e0}, \href
  {https://ui.adsabs.harvard.edu/abs/2018ApJ...856...74K} {856, 74}

\bibitem[\protect\citeauthoryear{{Kulkarni}, {Vogel}, {Wang}  \&
  {Wood}}{{Kulkarni} et~al.}{1992}]{Kulkarni1992}
{Kulkarni} S.~R.,  {Vogel} S.~N.,  {Wang} Z.,   {Wood} D.~O.~S.,  1992, \mn@doi
  [\nat] {10.1038/360139a0}, \href
  {https://ui.adsabs.harvard.edu/abs/1992Natur.360..139K} {360, 139}

\bibitem[\protect\citeauthoryear{{Leitherer}, {Chapman}  \&
  {Koribalski}}{{Leitherer} et~al.}{1995}]{Leitherer1995}
{Leitherer} C.,  {Chapman} J.~M.,   {Koribalski} B.,  1995, \mn@doi [ApJ]
  {10.1086/176140}, \href {http://adsabs.harvard.edu/abs/1995ApJ...450..289L}
  {450, 289}

\bibitem[\protect\citeauthoryear{{Lockman}}{{Lockman}}{1989}]{Lockman1989}
{Lockman} F.~J.,  1989, \mn@doi [\apjs] {10.1086/191383}, \href
  {https://ui.adsabs.harvard.edu/abs/1989ApJS...71..469L} {71, 469}

\bibitem[\protect\citeauthoryear{{Mackey}, {Green}  \& {Moutzouri}}{{Mackey}
  et~al.}{2020}]{Mackey2020}
{Mackey} J.,  {Green} S.,   {Moutzouri} M.,  2020, in Journal of Physics
  Conference Series. p. 012012 (\mn@eprint {arXiv} {2007.15357}),
  \mn@doi{10.1088/1742-6596/1620/1/012012}

\bibitem[\protect\citeauthoryear{{Ma{\'\i}z Apell{\'a}niz}, {Crespo Bellido},
  {Barb{\'a}}, {Fern{\'a}ndez Aranda}  \& {Sota}}{{Ma{\'\i}z Apell{\'a}niz}
  et~al.}{2020}]{Maiz2020}
{Ma{\'\i}z Apell{\'a}niz} J.,  {Crespo Bellido} P.,  {Barb{\'a}} R.~H.,
  {Fern{\'a}ndez Aranda} R.,   {Sota} A.,  2020, \mn@doi [\aap]
  {10.1051/0004-6361/202038228}, \href
  {https://ui.adsabs.harvard.edu/abs/2020A&A...643A.138M} {643, A138}

\bibitem[\protect\citeauthoryear{{Meyer}, {van Marle}, {Kuiper}  \&
  {Kley}}{{Meyer} et~al.}{2016}]{Meyer2016}
{Meyer} D.~M.-A.,  {van Marle} A.-J.,  {Kuiper} R.,   {Kley} W.,  2016, \mn@doi
  [\mnras] {10.1093/mnras/stw651}, \href
  {http://adsabs.harvard.edu/abs/2016MNRAS.459.1146M} {459, 1146}

\bibitem[\protect\citeauthoryear{{Meyer}, {Mignone}, {Kuiper}, {Raga}  \&
  {Kley}}{{Meyer} et~al.}{2017}]{Meyer2017}
{Meyer} D.~M.~A.,  {Mignone} A.,  {Kuiper} R.,  {Raga} A.~C.,   {Kley} W.,
  2017, \mn@doi [\mnras] {10.1093/mnras/stw2537}, \href
  {https://ui.adsabs.harvard.edu/abs/2017MNRAS.464.3229M} {464, 3229}

\bibitem[\protect\citeauthoryear{{Muijres}, {Vink}, {de Koter}, {M{\"u}ller}
  \& {Langer}}{{Muijres} et~al.}{2012}]{Muijres2012}
{Muijres} L.~E.,  {Vink} J.~S.,  {de Koter} A.,  {M{\"u}ller} P.~E.,   {Langer}
  N.,  2012, \mn@doi [A\&A] {10.1051/0004-6361/201015818}, \href
  {http://adsabs.harvard.edu/abs/2012A%26A...537A..37M} {537, A37}

\bibitem[\protect\citeauthoryear{{Ng}, {Bucciantini}, {Gaensler}, {Camilo},
  {Chatterjee}  \& {Bouchard}}{{Ng} et~al.}{2012}]{Ng2012}
{Ng} C.~Y.,  {Bucciantini} N.,  {Gaensler} B.~M.,  {Camilo} F.,  {Chatterjee}
  S.,   {Bouchard} A.,  2012, \mn@doi [\apj] {10.1088/0004-637X/746/1/105},
  \href {https://ui.adsabs.harvard.edu/abs/2012ApJ...746..105N} {746, 105}

\bibitem[\protect\citeauthoryear{{Noriega-Crespo}, {van Buren}  \&
  {Dgani}}{{Noriega-Crespo} et~al.}{1997}]{Noriega1997}
{Noriega-Crespo} A.,  {van Buren} D.,   {Dgani} R.,  1997, \mn@doi [\aj]
  {10.1086/118298}, \href
  {https://ui.adsabs.harvard.edu/abs/1997AJ....113..780N} {113, 780}

\bibitem[\protect\citeauthoryear{{Peri}, {Benaglia}, {Brookes}, {Stevens}  \&
  {Isequilla}}{{Peri} et~al.}{2012}]{Peri2012}
{Peri} C.~S.,  {Benaglia} P.,  {Brookes} D.~P.,  {Stevens} I.~R.,   {Isequilla}
  N.~L.,  2012, \mn@doi [\aap] {10.1051/0004-6361/201118116}, \href
  {https://ui.adsabs.harvard.edu/abs/2012A&A...538A.108P} {538, A108}

\bibitem[\protect\citeauthoryear{{Peri}, {Benaglia}  \& {Isequilla}}{{Peri}
  et~al.}{2015}]{Peri2015}
{Peri} C.~S.,  {Benaglia} P.,   {Isequilla} N.~L.,  2015, \mn@doi [\aap]
  {10.1051/0004-6361/201424676}, \href
  {https://ui.adsabs.harvard.edu/abs/2015A&A...578A..45P} {578, A45}

\bibitem[\protect\citeauthoryear{{Perley} \& {Butler}}{{Perley} \&
  {Butler}}{2013}]{Perley2013}
{Perley} R.~A.,  {Butler} B.~J.,  2013, \mn@doi [\apjs]
  {10.1088/0067-0049/206/2/16}, \href
  {https://ui.adsabs.harvard.edu/abs/2013ApJS..206...16P} {206, 16}

\bibitem[\protect\citeauthoryear{{Perley} \& {Butler}}{{Perley} \&
  {Butler}}{2017}]{Perley2017}
{Perley} R.~A.,  {Butler} B.~J.,  2017, \mn@doi [\apjs]
  {10.3847/1538-4365/aa6df9}, \href
  {https://ui.adsabs.harvard.edu/abs/2017ApJS..230....7P} {230, 7}

\bibitem[\protect\citeauthoryear{{Reich} \& {Schlickeiser}}{{Reich} \&
  {Schlickeiser}}{1992}]{Reich1992}
{Reich} W.,  {Schlickeiser} R.,  1992, \aap, \href
  {https://ui.adsabs.harvard.edu/abs/1992A&A...256..408R} {256, 408}

\bibitem[\protect\citeauthoryear{{S{\'a}nchez-Ayaso}, {del Valle},
  {Mart{\'\i}}, {Romero}  \& {Luque-Escamilla}}{{S{\'a}nchez-Ayaso}
  et~al.}{2018}]{Sanchez-Ayaso2018}
{S{\'a}nchez-Ayaso} E.,  {del Valle} M.~V.,  {Mart{\'\i}} J.,  {Romero} G.~E.,
   {Luque-Escamilla} P.~L.,  2018, \mn@doi [\apj] {10.3847/1538-4357/aac7c7},
  \href {https://ui.adsabs.harvard.edu/abs/2018ApJ...861...32S} {861, 32}

\bibitem[\protect\citeauthoryear{{Schulz}, {Ackermann}, {Buehler}, {Mayer}  \&
  {Klepser}}{{Schulz} et~al.}{2014}]{Schulz2014}
{Schulz} A.,  {Ackermann} M.,  {Buehler} R.,  {Mayer} M.,   {Klepser} S.,
  2014, \mn@doi [\aap] {10.1051/0004-6361/201423468}, \href
  {https://ui.adsabs.harvard.edu/abs/2014A&A...565A..95S} {565, A95}

\bibitem[\protect\citeauthoryear{{Taylor}, {Goss}, {Coleman}, {van Leeuwen}  \&
  {Wallace}}{{Taylor} et~al.}{1996}]{Taylor1996}
{Taylor} A.~R.,  {Goss} W.~M.,  {Coleman} P.~H.,  {van Leeuwen} J.,   {Wallace}
  B.~J.,  1996, \mn@doi [\apjs] {10.1086/192363}, \href
  {https://ui.adsabs.harvard.edu/abs/1996ApJS..107..239T} {107, 239}

\bibitem[\protect\citeauthoryear{{Taylor} et~al.,}{{Taylor}
  et~al.}{2003}]{CGPS2003}
{Taylor} A.~R.,  et~al., 2003, \mn@doi [\aj] {10.1086/375301}, \href
  {https://ui.adsabs.harvard.edu/abs/2003AJ....125.3145T} {125, 3145}

\bibitem[\protect\citeauthoryear{{Toal{\'a}}, {Oskinova},
  {Gonz{\'a}lez-Gal{\'a}n}, {Guerrero}, {Ignace}  \& {Pohl}}{{Toal{\'a}}
  et~al.}{2016}]{Toala2016}
{Toal{\'a}} J.~A.,  {Oskinova} L.~M.,  {Gonz{\'a}lez-Gal{\'a}n} A.,  {Guerrero}
  M.~A.,  {Ignace} R.,   {Pohl} M.,  2016, \mn@doi [\apj]
  {10.3847/0004-637X/821/2/79}, \href
  {https://ui.adsabs.harvard.edu/abs/2016ApJ...821...79T} {821, 79}

\bibitem[\protect\citeauthoryear{{Toal{\'a}}, {Oskinova}  \&
  {Ignace}}{{Toal{\'a}} et~al.}{2017}]{Toala2017}
{Toal{\'a}} J.~A.,  {Oskinova} L.~M.,   {Ignace} R.,  2017, \mn@doi [\apjl]
  {10.3847/2041-8213/aa667c}, \href
  {https://ui.adsabs.harvard.edu/abs/2017ApJ...838L..19T} {838, L19}

\bibitem[\protect\citeauthoryear{{Uyaniker}, {Furst}, {Reich}, {Aschenbach}  \&
  {Wielebinski}}{{Uyaniker} et~al.}{2001}]{Uyaniker2001}
{Uyaniker} B.,  {Furst} E.,  {Reich} W.,  {Aschenbach} B.,   {Wielebinski} R.,
  2001, \mn@doi [\aap] {10.1051/0004-6361:20010387}, \href
  {https://ui.adsabs.harvard.edu/abs/2018ApJ...856...74K} {371, 675}

\bibitem[\protect\citeauthoryear{{Wilkin}}{{Wilkin}}{1996}]{Wilkin1996}
{Wilkin} F.~P.,  1996, \mn@doi [\apjl] {10.1086/309939}, \href
  {http://adsabs.harvard.edu/abs/1996ApJ...459L..31W} {459, L31}

\bibitem[\protect\citeauthoryear{{Yusef-Zadeh} \& {Gaensler}}{{Yusef-Zadeh} \&
  {Gaensler}}{2005}]{Yusef-Zadeh2005}
{Yusef-Zadeh} F.,  {Gaensler} B.~M.,  2005, \mn@doi [Advances in Space
  Research] {10.1016/j.asr.2005.03.003}, \href
  {https://ui.adsabs.harvard.edu/abs/2005AdSpR..35.1129Y} {35, 1129}

\bibitem[\protect\citeauthoryear{{del Palacio}, {Bosch-Ramon}, {M{\"u}ller}  \&
  {Romero}}{{del Palacio} et~al.}{2018}]{delPalacio2018}
{del Palacio} S.,  {Bosch-Ramon} V.,  {M{\"u}ller} A.~L.,   {Romero} G.~E.,
  2018, \mn@doi [\aap] {10.1051/0004-6361/201833321}, \href
  {https://ui.adsabs.harvard.edu/abs/2018A&A...617A..13D} {617, A13}

\bibitem[\protect\citeauthoryear{{del Valle} \& {Pohl}}{{del Valle} \&
  {Pohl}}{2018}]{DelValle2018}
{del Valle} M.~V.,  {Pohl} M.,  2018, \mn@doi [\apj]
  {10.3847/1538-4357/aad333}, \href
  {https://ui.adsabs.harvard.edu/abs/2018ApJ...864...19D} {864, 19}

\bibitem[\protect\citeauthoryear{{del Valle} \& {Romero}}{{del Valle} \&
  {Romero}}{2012}]{delValle2012}
{del Valle} M.~V.,  {Romero} G.~E.,  2012, \mn@doi [\aap]
  {10.1051/0004-6361/201218937}, \href
  {https://ui.adsabs.harvard.edu/abs/2012A&A...543A..56D} {543, A56}

\bibitem[\protect\citeauthoryear{{del Valle} \& {Romero}}{{del Valle} \&
  {Romero}}{2014}]{delValle2014}
{del Valle} M.~V.,  {Romero} G.~E.,  2014, \mn@doi [\aap]
  {10.1051/0004-6361/201322308}, \href
  {http://adsabs.harvard.edu/abs/2014A%26A...563A..96D} {563, A96}

\bibitem[\protect\citeauthoryear{{van Marle}, {Decin}  \& {Meliani}}{{van
  Marle} et~al.}{2014}]{vanmarle2014}
{van Marle} A.~J.,  {Decin} L.,   {Meliani} Z.,  2014, \mn@doi [\aap]
  {10.1051/0004-6361/201321968}, \href
  {https://ui.adsabs.harvard.edu/abs/2014A&A...561A.152V} {561, A152}

\makeatother
\end{thebibliography}

%%%%%%%%%%%%%%%%%%%%%%%%%%%%%%%%%%%%%%%%%%%%%%%%%%

%%%%%%%%%%%%%%%%% APPENDICES %%%%%%%%%%%%%%%%%%%%%
\end{document}